\documentclass[a4paper, 11pt]{article}
\usepackage{}
\usepackage{mathrsfs}
\usepackage{amssymb}
\usepackage{bbm}
\usepackage{epsfig}
\usepackage{amsmath}
\usepackage{appendix}
\usepackage[english]{babel}

\def\Reals{\mathop{\hbox{\mit I\kern-.2em R}}\nolimits}

\def\Complexes{{\hbox{\mit C\kern-.46em
            \vrule depth 0ex height 1.4ex width .05em\kern.41em}}}

\newtheorem{theorem}{Theorem}[section]
\newtheorem{defn}{Definition}[section]
\newtheorem{coro}{Corollary}[section]
\newtheorem{lem}{Lemma}[section]
\newtheorem{remark}{Remark}[section]

\title{\bf Connectivity and Set Tracking of Multi-agent Systems Guided by Multiple Moving
Leaders\footnote{This work has been supported in part
by the NNSF of China under Grants 60874018, 60736022, and
60821091, the Knut and Alice Wallenberg Foundation and the Swedish Research Council.}}
\date{}

\author{Guodong Shi, Yiguang Hong\thanks{G. Shi and Y. Hong are with Key Laboratory of Systems and Control, Institute of Systems
       Science, Chinese Academy of Sciences, Beijing 100190, China.
       Email: {\tt\small shigd@amss.ac.cn, yghong@iss.ac.cn}} and Karl Henrik Johansson
\thanks{K. Johansson is with ACCESS Linnaeus Centre, School of Electrical Engineering,
Royal Institute of Technology, Stockholm 10044, Sweden.
       Email: {\tt\small kallej@ee.kth.se}}
}

\begin{document}
\maketitle
\begin{abstract}
In this paper, we investigate distributed multi-agent tracking
of a convex set specified by multiple moving leaders with
unmeasurable velocities. Various jointly-connected interaction
topologies of the follower agents with uncertainties are considered
in the study of set tracking. Based on the connectivity of the
time-varying multi-agent system, necessary and sufficient conditions
are obtained for set input-to-state stability and set
integral input-to-state stability for a nonlinear
neighbor-based coordination rule with switching directed topologies.
Conditions for asymptotic set tracking are also proposed with respect to
the polytope spanned by the leaders.
\end{abstract}


\section{Introduction}

The last decade has witnessed tremendous interest devoted to the
investigation of collective phenomena in multiple autonomous agents,
due to broad applications in various fields of science ranging from
biology to physics, engineering, and ecology, just to name a few
\cite{iain, fang, mar, ren, sabertac}. Concerning the issues of
multi-agent systems and distributed design, the revolutionary idea
is underpinning a strong interaction of individual dynamics,
communication topologies, and distributed controls. The problem is
generally very challenging due to the complex dynamics and
hierarchical structures of the systems. However, efforts have been
started with relatively simple problems such as consensus,
formation, and rendezvous, and many significant results have been
obtained.

The leader-follower coordination is an important multi-agent control
problem, where the leader may be a real leader (such as a target, an
evader, or a predefined position), or a virtual leader (such as a
reference trajectory or a specified path). In most theoretical
work, a single leader with exact measurement is considered on
multi-agent systems for each agent to follow. However, in practical
situations, multiple leaders and target sets with unmeasurable
variables are considered to achieve desired collective
behaviors. In \cite{iain}, a simple model was given to simulate fish
foraging and demonstrate the leader effectiveness when the leaders
(or informed agents) guide a school of fish to a particular food
region. In \cite{lin05}, a straight-line formation of a group of
agents was discussed, where all the agents converge to the line
segment specified by two edge leaders. A containment
control scheme was proposed with fixed undirected interaction in
\cite{ji}, which aimed at driving a group of agents to a given
target location and making their positions contained in the polytope
spanned by multiple stationary or moving leaders during their motion.
Region following formation control was constructed \cite{che}, where all
the robots are driven and then stay within a moving target region as
a group. Moreover, different dynamic connectivity
conditions were obtained to guarantee that the multiple leaders (or
informed agents) aggregate the whole multi-agent group within a
convex target set in \cite{shi09}. Additionally, control strategies
were demonstrated and analyzed to drive a collection of mobile
agents to stationary/moving leaders with connectivity-maintenance
and collision-avoidance with fixed and switching directed network
topologies in \cite{caoren}. As a matter of fact, multiple leaders
are usually assigned to increase control effectiveness, enhance
communication/sensing range, improve reliability, and optimize
energy cost in multi-agent coordination.

Connectivity plays a key role in the coordination of multi-agent
networks, which is related to the influence of agents and
controllability of the network. Due to mobility of the agents,
inter-agent topologies usually keep changing in practice. Therefore,
the various connectivity conditions to describe frequently switching
topologies in order to deal with multi-agent consensus or flocking
\cite{saber04, hong, tan, xiao}.  In fact, the ``joint connection"
or similar concepts are important  in the analysis of stability and
convergence to guarantee multi-agent coordination with
time-dependent topology. Uniformly jointly-connected conditions have
been employed for different problems. \cite{tsi} studied the
distributed asynchronous iterations, while \cite{jad03} proved the
consensus of a simplified Vicsek model.  Furthermore, \cite{hong07}
and \cite{cheng} investigated the jointly-connected coordination for
second-order agent dynamics via different approaches, while
\cite{lin07} worked on nonlinear continuous-time agent dynamics with
jointly-connected interaction graphs. Also, flocking of multi-agent
system with state-dependent topology was studied with non-smooth
analysis in \cite{tan, tantac}. What is more, the $[t, \infty)$
joint connection condition, which is more generalized than the
uniformly joint connection assumption, was discussed by Moreau, in
order to achieve the consensus for discrete-time agents in
\cite{mor}. This $[t, \infty)$ connectivity concept was then
extended in the distributed control analysis for the target set
convergence in \cite{shi09}.

It is well known that input-to-state stability (ISS) is an important
and very useful tool in the study of the stability and stabilization
of control systems \cite{jiang,sontag-wang}.  Its variants such
as integral input-to-state stability (iISS) were discussed in
\cite{sontag1}. Then few works on set input-to-state stability
(SISS) were done with respect to fixed sets in \cite{Sontag}. On the
other hand, ISS or related ideas can facilitate the control
analysis and synthesis with interconnection conditions like small
gains (referring to \cite{jiang}, for example). ISS has recently been
applied to the stability study of a group of interconnected
nonlinear systems \cite{cdc09}. Moreover, an extended concept called
leader-to-formation stability was introduced to investigate the
stability of the formation of a group of agents in light of ISS
properties \cite{tan04}.  In fact, ISS application in multi-agent
systems is promising.

The contributions of the paper include:
\begin{itemize}
\item We propose
the generalized set input-to-state stability (SISS) and set
integral-input-to-state stability (SiISS) to handle moving sets with
time-varying shapes for switching multi-agent networks.

\item We study the multi-leader coordination from the ISS viewpoint.
With the help of SISS and SiISS, we give explicit expressions to
estimate the convergence rate and tracking error of a group of
mobile agents that try to enter the convex hull determined by
multiple leaders.

\item We show relationships between the connectivity
and set tracking of the multi-agent system, and find that various
jointly-connected conditions usually provide necessary and/or
sufficient conditions for distributed coordination.

\item We develop a method to study SISS and SiISS for a moving
set and switching topology with graph theory and non-smooth
analysis. In fact, we cannot take the standard approaches to
conventional ISS or iISS using equivalent ISS-Lyapunov functions
\cite{sontag1,sontag-wang}. In addition, the classic algebraic
methods based on Laplacian may fail due to disturbances in nonlinear
agent dynamics, uncertain leader velocities, or moving multi-leader
set.
\end{itemize}

This paper is organized as follows.  Section 2 presents the
preliminaries and problem formulation, while Section 3 proposes
results for the convergence estimation. Section 4 mainly reports a
necessary and sufficient condition for the SISS with respect to the
moving multi-leader set with switching inter-agent topologies, and
then presents a set-tracking case based on the SISS. Correspondingly,
Section 5 obtains necessary and sufficient conditions for SiISS
and then shows set-tracking results related to SiISS.  Finally,
Section 6 gives concluding remarks.

\section{Problem Formulation}

In this section, we introduce some preliminary knowledge for the
following discussion.

First we introduce some basic concepts in graph theory (referring to
\cite{god} for details). A directed graph (digraph) $\mathcal
{G}=(\mathcal {V}, \mathcal {E})$ consists of a finite set
$\mathcal{V}=\{1,2,...,n\}$ of nodes and an arc set
$\mathcal {E}$, in which an arc is an ordered pair of
distinct nodes of $\mathcal {V}$.  $(i,j)\in\mathcal {E}$ describes
an arc which leaves $i$ and enters $j$.  A {\it walk} in digraph $\mathcal
{G}$ is an alternating sequence $\mathcal
{W}:
i_{1}e_{1}i_{2}e_{2}\dots e_{m-1}i_{m}$ of nodes $i_{\kappa}$ and
arcs $e_{\kappa}=(i_{\kappa},i_{\kappa+1})\in\mathcal {E}$ for
$\kappa=1,2,\dots,m-1$. A walk  is called a {\it path}
if the nodes of this walk are distinct, and a path from $i$ to
$j$ is denoted as $\widehat{(i,j)}$. Node $j$ is called {\it reachable}
from $i$ if there is a path $\widehat{(i,j)}$. If the nodes
$i_1,\dots,i_{m-1}$ are distinct and $i_1=i_m$, $\mathcal
{W}$ is called a (directed) {\it cycle}. A digraph without cycles is
said to be {\it acyclic}.

The union of the two digraphs $\mathcal {G}_1=(\mathcal {V},\mathcal {E}_1)$ and $\mathcal
{G}_2=(\mathcal {V},\mathcal {E}_2)$ is defined as $\mathcal {G}_1\cup\mathcal
{G}_2=(\mathcal {V},\mathcal {E}_1\cup\mathcal {E}_2)$ if they have the same node set. Furthermore, a
time-varying digraph is defined as $\mathcal
{G}_{\sigma(t)}=(\mathcal {V},\mathcal {E}_{\sigma(t)})$ with
$\sigma:t\rightarrow \mathcal {Q}$ as a piecewise constant function,
where $\mathcal {Q}$ is the finite set which consists of all the possible digraphs
with node set $\mathcal {V}$. Moreover, the joint digraph of $\mathcal
{G}_{\sigma(t)}$ in
time interval $[t_1,t_2)$ with $t_1<t_2\leq +\infty$ is denoted as
\begin{equation}
\label{jointgraph} \mathcal {G}([t_1,t_2))= \cup_{t\in[t_1,t_2)}
\mathcal {G}(t)=(\mathcal {V},\cup_{t\in[t_1,t_2)}\mathcal
{E}_{\sigma(t)}).
\end{equation}

Next, we recall some notations in convex analysis (see \cite{rock}).
A set $K\subset R^d$ is said to be convex if $(1-\lambda)x+\lambda
y\in K$ whenever $x\in K,y\in K$ and $0\leq\lambda \leq1$.
%A vector sum $\lambda_1 x_1+\dots+\lambda_n x_n$ is called a
%convex combination of $x_1,\dots,x_n$ if the coefficients
%$\lambda_i$ are all non-negative and $\lambda_1+\dots+\lambda_n=1$.
For any set $S\subset R^d$, the intersection of all convex sets
containing $S$ is called the {\it convex hull} of $S$, denoted by
$co(S)$. Particularly, the convex hull of a finite set of points
$x_1,\dots,x_n\in R^d$ is a polytope, denoted by
$co\{x_1,\dots,x_n\}$. In fact, we have
$co\{x_1,\dots,x_n\}=\{\lambda_1 x_1+\dots+\lambda_n
x_n|\lambda_1+\dots+\lambda_n=1,\lambda_i\geq 0\}$.

Let $K$ be a closed convex subset in $R^d$ and denote
$|x|_K\triangleq\inf\{| x-y | \mid y\in K\}$, where $|\cdot|$
denotes the Euclidean norm for a vector or the absolute value of a
scalar (\cite{sontag-wang, sontag1}). Then we can associate to any
$x\in R^d$ a unique element $\mathcal{P}_{K}(x)\in K$ satisfying
$|x-\mathcal{P}_{K}(x)|=|x|_K,$
where the map $\mathcal{P}_{K}$ is called the projector onto $K$ and
\begin{equation}\label{r9}
\langle \mathcal{P}_{K}(x)-x,\mathcal{P}_{K}(x)-y\rangle\leq 0,\quad \forall y\in
K.
\end{equation}
Clearly, $|x|_K^2$ is continuously differentiable at point $x$, and (see \cite{aubin})
\begin{equation}\label{r10}
\nabla |x|_K^2=2(x-\mathcal{P}_{K}(x)).
\end{equation}

The following lemma was obtained in \cite{shi09}, which is useful in
what follows.

\begin{lem}
\label{lem6} Suppose $K\subset R^d$ is a convex set and
$x_a,x_b\in R^d$. Then
\begin{equation}\label{9}
\langle x_a-\mathcal {P}_K(x_a),x_b-x_a\rangle\leq
|x_a|_K\cdot     \left.| |x_a|_K-|x_b|_K \right|.
\end{equation}
Particularly, if $|x_a|_K>|x_b|_K$, then
\begin{equation}
\langle x_a-\mathcal {P}_K(x_a),x_b-x_a\rangle\leq -|x_a|_K\cdot(
|x_a|_K-|x_b|_K).
\end{equation}
\end{lem}

Then we consider the Dini derivative for the following non-smooth
analysis. Let $a$ and $b\;(>a)$ be two real numbers and consider a
function $h: (a,b)\to R$ and a point $t\in (a,b)$. The upper Dini
derivative of $h$ at $t$ is defined as
$$
D^+h(t)=\limsup_{s\to 0^+} \frac{h(t+s)-h(t)}{s}.
$$
It is well known that  when $h$ is continuous on $(a,b)$, $h$ is
non-increasing on $(a,b)$ if and only if $ D^+h(t)\leq 0$ for any
$t\in (a,b)$ (more details can be found in \cite{rou}). The next
result is given for the calculation of Dini derivative
\cite{dan,lin07}.

\begin{lem}
\label{lem3}  Let $V_i(t,x): R\times R^d \to R\;(i=1,\dots,n)$ be
$C^1$ and $V(t,x)=\max_{i=1,\dots,n}V_i(t,x)$. If $
\mathcal{I}(t)=\{i\in \{1,2,\dots,n\}\,:\,V(t,x(t))=V_i(t,x(t))\}$
is the set of indices where the maximum is reached at $t$, then
$
D^+V(t,x(t))=\max_{i\in\mathcal{ I}(t)}\dot{V}_i(t,x(t)).
$
\end{lem}

\begin{figure}
\centerline{\epsfig{figure=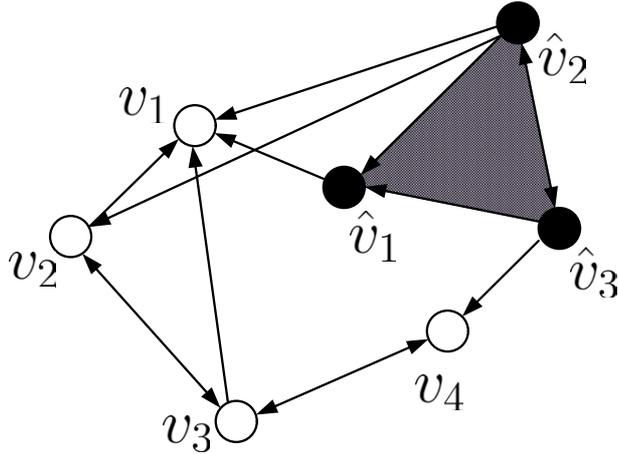, width=0.5\linewidth=0.25}}
\caption{Multiple agents ($v_i, \, i=1,2,3,4$) with multiple leaders
($\hat v_i, \, i=1,2,3$)}\label{fig0}
\end{figure}

In this paper, we consider the set coordination problems for a
multi-agent system consisting of $n$ follower-agents and $k$ leader-agents
(see Fig. \ref{fig0}).  The follower set is denoted as
$\mathcal{V}_{F}\triangleq\{v_1,\dots,v_n\}$, and the leader set is
denoted as
$\mathcal{V}_{L}\triangleq\{\hat{v}_{1},\dots,\hat{v}_{k}\}$. In
what follows, we will identify follower $v_i$ or leader $\hat{v}_i$
with its index $i$ (namely, agent $i$ or leader $i$) if there is no
confusion.

Then we describe the communication in the multi-agent network. At
time $t$, if $i\in\mathcal {V}_F$ can ``see" $j\in\mathcal
{V}_F$, there is an arc $(j,i)$ (marking the information flow) from
 $j$ to $i$, and then agent $j$ is said to be a {\it neighbor} of
agent $i$. Moreover, if $i\in\mathcal {V}_F$ ``sees"
$j\in\mathcal {V}_L$ at time $t$, there is an arc $(j,i)$ leaving
from $j$ and entering $i$, and then $j$ is said to be a {\it leader}
of agent $i$. Let $N_i$ and $L_i$ represent the set of agent $i$'s
neighbors and the set of agent $i$'s leaders (that is, the leaders which
are connected to agent $i$), respectively. Note that, since the
leaders are not influenced by the followers, there is no arc leaving
from $\mathcal {V}_F$ entering $\mathcal {V}_L$.

Define $\mathcal {V}=\mathcal{V}_{F}\cup\mathcal{V}_{L}$ as the
whole agent set (including leaders and followers). Denote $\mathcal
{P}$ as the set of all possible interconnection topologies, and
$\sigma: [0,+\infty)\rightarrow \mathcal {P}$ as a piecewise
constant switching signal function to describe the switchings
between the topologies. Thus, the interaction topology of the
considered multi-agent network is described by a time-varying
directed graph $\mathcal {G}_{\sigma(t)}=(\mathcal {V},\mathcal
{E}_{\sigma(t)})$. Correspondingly, $\mathcal {G}^F_{\sigma(t)}$ is denoted as the communication graph among the follower agents. Additionally, let $N_i(\sigma(t))$ and
$L_i(\sigma(t))$ represent the set of agent $i$'s neighbors and the
set of its connected leaders in $\mathcal {G}_{\sigma(t)}$,
respectively.

As usual in the literature \cite{jad03,lin07,shi09}, an assumption is given for the switching signal $\sigma(t)$.

\noindent {\bf Assumption 1} (Dwell Time) There is a
 lower bound $\tau_D>0$ between two
switching instants.

We give definitions for the connectivity of a multi-agent system
with multiple leaders.

\begin{defn}
(i) $\mathcal {G}_{\sigma(t)}$ is said to be {\it L-connected} if, for
any $i\in\mathcal{V}_{F}$, there exists a leader
$j\in\mathcal{V}_{L}$ such that there is a path from leader $j$ to
agent $i$ in $\mathcal {G}_{\sigma(t)}$ at time $t$. Moreover, $\mathcal {G}_{\sigma(t)}$ is said to be {\it jointly
L-connected} in time interval $[t_1,t_2)$ if the union graph
$\mathcal {G}([t_1,t_2))$ is L-connected;

(ii) $\mathcal {G}_{\sigma(t)}$ is said to be {\it jointly
L-connected} (JLC) if the union graph $\mathcal {G}([t,\infty))$ is
L-connected for any $t$;

(iii) $\mathcal {G}_{\sigma(t)}$ is said to be {\it uniformly
jointly L-connected} (UJLC) if there exists $T>0$ such that the
union graph $\mathcal {G}([t,t+T))$ is L-connected for any $t\geq
0$.
\end{defn}

\begin{remark}
Note that the L-connectedness describes the capacity for the
follower agents to get the information from the moving multi-leader
set in the information flow, and an L-connected graph may not be
connected since the graph with leaders as its nodes may not be
connected. In fact, if we consider the group of the leaders as one
virtual node in $\mathcal {V}$, then the L-connectedness becomes the
quasi-strong connectedness for a digraph \cite{ber,lin07}.
\end{remark}

The state of agent $v_i\in \mathcal{V}_{F}$, is denoted as $x_i \in
R^d\,(i=1,\dots,n$), and the state of leader $\hat{v}_i\in
\mathcal{V}_{L}$, is denoted as $y_i \in R^d\,(i=1,\dots,k$).
Denote $x=(x_1,\dots,x_n)^T\in R^{nd}$ and
$y=(y_{1},\dots,y_{k})^T\in R^{kd}$ and let the continuous function
$a_{ij}(x,y, t)>0$ be the weight of arc $(j,i)$, if any, for $i,j\in
\mathcal {V}_F$, and continuous function $b_{ij}(x,y, t)>0$ be the
weight of arc $(j,i)$, if any, for $i\in\mathcal {V}_F;j\in\mathcal
{V}_L$.

Then we present the multi-agent model for the active leaders and the
(follower) agents
\begin{equation}\label{5}
\left\{
\begin{array}{ll}
\dot{y}_{i}=u_i(y,t), \quad \quad i=1,\dots,k\\
\dot{x}_{i}=\sum\limits_{j \in
N_i(\sigma(t))}a_{ij}(x,y,t)(x_j-x_i)+\sum\limits_{j \in
L_i(\sigma(t))}b_{ij}(x,y,t)(y_j-x_i)+w_i(t), \;\quad i=1,\dots,n
\end{array}
\right.
\end{equation}
where $u_i(y,t)$ describes the control inputs of the leader
$i,i\in\mathcal {V}_L$, which is continuous in $y$ for fixed $t$ and
piecewise continuous in $t$ for fixed $y$, and $w_i(t)$ is a
continuous function to describe the disturbances in communication
links and individual dynamics to follower agent $i$. Then another assumption is given on the weight functions $a_{ij}(x,y,t)$ and
$b_{ij}(x,y,t)$.

\noindent {\bf Assumption 2} (Bounded Weights) There are $0<a_\ast\leq a^\ast$ and $b_\ast>0$ such that
$a_\ast\leq a_{ij}(x,y, t)\leq a^\ast,\; b_\ast\leq b_{ij}(x,y, t)$
for any $x,y, t$.

\begin{remark}
In (\ref{5}), the weights, $a_{ij}$ and $b_{ij}$, may not be
constant. Instead, because of the complex communication and
environment uncertainties, they are dependent on time or space or
relative measurement (see nonlinear models given in \cite{lin07,
shi09, mor, tan}). Some models such as those studied in \cite{lin07,
shi09} can be written in the form of (\ref{5}), while other
nonlinear multi-agent models may be transformed to this class of
multi-agent systems in some situations.  Here $a_{ij}(x,y,t)$ and
$b_{ij}(x,y,t)$ are written in a general form simply for
convenience, and global information is not required in our study.
For example, $a_{ij}$ and $b_{ij}$ can depend only on the state of
$x_i$, time $t$ and $x_j\, (j\in N_i)$, which is certainly a special
form of $a_{ij}(x,y,t)$ or $b_{ij}(x,y,t)$. In other words, the
control laws in specific decentralized forms are still
decentralized.
\end{remark}

Without loss of generality, we assume the initial time $t=0$, and
the initial condition $x^0=(x_1(0),\dots,x_n(0))^T\in R^{nd}$ and
$y^0=(y_1(0),\dots,y_k(0))^T\in R^{kd}$.

Denote the time-varying polytope formed by the $k$ active leaders
\begin{equation}
\label{movset} \mathcal {L}(y(t))\triangleq
co\{y_1(t),\dots,y_k(t)\},
\end{equation}
and let
$$
|x(t)|_{\mathcal {L}(y(t))}\triangleq \max_{i\in\mathcal {V}_F} |x_i(t)|_{\mathcal {L}(y(t))}
$$
be the maximal distance for the followers away from the moving
multi-leader set $\mathcal {L}(y(t))$.

The following definition is to describe the convergence to the
moving convex set $\mathcal {L}(y(t))$.

\begin{defn}
\label{def-set} The (global) {\it set tracking} (ST) with respect to
${\cal L}(y(t))$ for system (\ref{5}) is achieved if
\begin{equation}
\lim_{t\rightarrow +\infty} |x(t)|_{\mathcal {L}(y(t))}=0
\end{equation}
for any initial condition $x^0\in R^{nd}$ and $y^0\in R^{kd}$.
\end{defn}

For a stationary convex set $K$, set tracking can be reduced to set
stability and attractivity, and methods to analyze $|x_i(t)|_{K}$
were proposed in some existing works \cite{shi09}.  In fact,
\cite{ji, caoren} discussed the convergence to the static convex set
determined by stationary leaders with well designed control
protocols. Moreover, if we assume that the target set is exactly the
polytope with the positions of the stationary leaders (or informed
agents) as its vertices, then the convergence to the polytope,
treated as a target set, can be obtained straightforwardly based on
the results and limit-set-based methods given in \cite{shi09}.

Input-to-state stability has been widely used in the stability
analysis and set input-to-state stability (SISS) for a fixed set has
been studied in \cite{Sontag}.  To study the multi-leader set
tracking in a broad sense, we introduce a generalized SISS with
respect to $\mathcal {L}(y(t))$, a moving set with a time-varying
shape, for multi-agent systems with switching interaction
topologies. Denote $u\triangleq(u_1,\dots,u_k)^T$,
$w\triangleq(w_1,\dots,w_n)^T$, $z\triangleq (u^T\; w^T)^T$, and
$L_{\infty}\triangleq \{z:R_{\geq 0}\rightarrow
R^{(n+k)m}\;|\,\|z\|_\infty<\infty\}$ with $\|z\|_\infty\triangleq
\sup\{|z(t)|,t\geq 0\}$ (\cite{sontag-wang}).

A function $\gamma: R_{\geq 0}\rightarrow R_{\geq 0}$ is said to be
a $\mathcal{K}$-class function if it is continuous, strictly
increasing, and $\gamma(0)=0$. Moreover, a function $ \beta:R_{\geq
0}\times R_{\geq 0}\rightarrow R $ is a $\mathcal {KL}$-class function
if $\beta(\cdot,t)$ is of class $\mathcal {K}$ for each fixed $t\geq
0$ and $\beta(s,t)$ decreases to $0$ as $t\rightarrow\infty$ for
each fixed $s\geq 0$.

\begin{defn}
\label{def-siss} System (\ref{5}) is said to be globally generalized
set input-to-state stable (SISS) with respect to ${\cal L}(y(t))$
with input $z$ if there exist a $\mathcal {K}\mathcal {L}$-function
$\beta$ and a $\mathcal {K}$-function $\gamma$ such that
\begin{equation}
\label{siss1} |x(t)|_{\mathcal {L}(y(t))}\leq
\beta(|x^0|_{\mathcal {L}(y^0)},t )+\gamma(\|z
\|_{\infty})
\end{equation}
for $z\in L_\infty$ and any initial conditions $x^0\in R^{nd}$ and
$y^0\in R^{kd}$.
\end{defn}

%\begin{remark}
%It is obvious that (\ref{siss1}) is equivalent to that there exist a
%$\mathcal {K}\mathcal {L}$-function $\beta:R$ and a $\mathcal {K}$-function $\tilde \gamma$ such
%that,
%\begin{equation}
%\label{siss2} |x(t)|_{\mathcal {L}(y(t))}\leq
%\beta(|x^0|_{\mathcal {L}(y^0)},t )+\tilde \gamma(\|
%(u^T\; w^T)^T
%\|_{\infty})
%\end{equation}
%for any initial condition $x^0\in R^{nd}$ and $y^0\in R^{kd}$, or
%equivalent to that there are a $\mathcal {K}\mathcal {L}$-function
%$\hat\beta:R_{\geq 0}\times R_{\geq 0}\rightarrow R$ and a $\mathcal
%{K}$-function $\hat \gamma$ such that
%\begin{equation}
%\label{siss3} |x(t)|_{\mathcal {L}(y(t))}\leq
%\gamma(\| (u^T\; w^T)^T \|_{\infty})\}
%\end{equation}
%for any initial condition.
%\end{remark}

Integral-input-to-state stability (iISS) was introduced as an
integral variant of ISS, which has been proved to be strictly weaker
than ISS \cite{sontag1}. We also introduce a definition of
(generalized) set integral-input-to-state stability (SiISS) with
respect to a time-varying and moving set.

\begin{defn}
\label{def-siiss} System (\ref{5}) is (globally) generalized set
integral-input-to-state stable (SiISS) with respect to ${\cal
L}(y(t))$ if there exist a $\mathcal {K}\mathcal
{L}$-function $\beta$ and a $\mathcal {K}$-function $\gamma$ such
that
\begin{equation}\label{r2}
 |x(t)|_{\mathcal {L}(y(t))}\leq
\beta(|x^0|_{\mathcal {L}(y^0)},t )+\int_0^t \gamma(|z(s)|)ds,
\end{equation}
 for any initial conditions $x^0\in R^{nd}$ and
$y^0\in R^{kd}$.
\end{defn}

The conventional SISS was given for a fixed set $K$ (\cite{Sontag}),
while the generalized SISS or SiISS is proposed with respect to a
time-varying set $\mathcal {L}(y(t))$. In the following, we still
use SISS or SiISS instead of generalized SISS or SiISS for
simplicity.

\begin{remark}
Similar to the study of conventional ISS, local SISS and SiISS can
be defined. In this paper, we focus on the global SISS and SiISS. In
fact, it is rather easy to extend research ideas of global set
tracking to study local cases.
\end{remark}

\section{Convergence Estimation}

For the set tracking with respect to a moving multi-leader set of
system (\ref{5}), we have to deal with the estimation of
$|x_i(t)|_{\mathcal {L}(y(t))}$ when $\mathcal {L}(y(t))$ is a
time-varying convex set, where $y(t)$ is a trajectory of the moving
leaders in system (\ref{5}) with initial condition $y^0=y(0)$.
Define
\begin{equation}
\label{rq} r(t)\triangleq \max_{i\in\mathcal
{V}_L}|u_i(y(t),t)|;\quad q(t)\triangleq \max_{i\in\mathcal
{V}_L}|u_i(y(t),t)|+\max_{i\in\mathcal
{V}_F}|w_i(t)|.
\end{equation}
Obviously,
\begin{equation}
\label{g2} q(t) \leq |u(y(t),t)|+|w(t)| \leq \sqrt{2}|z(t)| \leq
\sqrt{2}\max\{\sqrt{n},\sqrt{k}\}q(t).
\end{equation}

The following result is given to estimate the changes of the
distance between an agent and the convex hull spanned by the
leaders.

\begin{lem}
\label{lem7}
For any $t, t_0\geq 0$ and $i=1,\dots,n$,
\begin{equation}\label{23}
||x_i(t)|_{\mathcal {L}(y(t))}-|x_i(t)|_{\mathcal {L}(y(t_0))}|\leq \int_{t_0}^{t}r(s)ds.
\end{equation}
\end{lem}

\noindent Proof: Suppose
$$
\mathcal {P}_{\mathcal {L}(y(t_0))}(x_i(t))=\sum_{i=1}^{k} \lambda_i
y_i(t_0)\in \mathcal {L}(y(t_0)),
$$
where $\lambda_i\geq 0$ for $i=1,\dots,k$ with $\sum\limits_{i=1}^k
\lambda_i=1$. Define $\hat y(t)\triangleq \sum\limits_{i=1}^{k}
\lambda_i y_i(t)$, and then
$$
|\hat y(t)-\hat y(t_0)|\leq \sum_{i=1}^{k} \lambda_i |
y_i(t)-y_i(t_0)| = \sum_{i=1}^{k} \lambda_i
|\int_{t_0}^{t}u_i(y(s),s)ds|\leq \int_{t_0}^{t}r(s)ds
$$
Moreover,
\begin{eqnarray}\label{101}
|x_i(t)|_{\mathcal {L}(y(t))} &\leq& |x_i(t)-\hat y(t)|\nonumber\\
&\leq& |x_i(t)-\hat y(t_0)|+|\hat y(t)-\hat y(t_0)|\nonumber\\
&\leq& |x_i(t)|_{\mathcal {L}(y(t_0))}+\int_{t_0}^{t}r(s)ds
\end{eqnarray}
Also, similar analysis leads to
\begin{equation}\label{102}
|x_i(t)|_{\mathcal {L}(y(t_0))}\leq |x_i(t)|_{\mathcal {L}(y(t))}+\int_{t_0}^{t}r(s)ds
\end{equation}
Therefore, (\ref{101}) and (\ref{102}) lead to the conclusion.
\hfill$\square$

For simplicity, define $ \psi_i(t)\triangleq |x_i(t)|^2_{\mathcal
{L}(y(t))},\;i=1,\dots,n $ and
$$
\Psi(t) \triangleq \max_{i\in \mathcal {V}_F}\psi_i(t),
$$
which is locally Lipschitz but may not be continuously
differentiable.  Clearly, $|x_i(t)|_{\mathcal
{L}(y(t))}=\sqrt{\psi_i(t)},i=1,\dots,n$ and $\|x(t)\|_{\mathcal
{L}(y(t))}=\sqrt{\Psi(t)}$.

Then, we get the following lemma to estimate the set convergence.

\begin{lem}
\label{lem10} $D^+\sqrt{\Psi(t)}\leq q(t)$.
\end{lem}

\noindent Proof: It is not hard to see that
\begin{eqnarray}\label{103}
\frac{d \psi_i(t)}{dt} &=&\lim_{\Delta t\rightarrow 0}\frac{\psi_i(t+\Delta t)-\psi_i(t)}{\Delta t}\nonumber\\
&=&\lim_{\Delta t\rightarrow 0}\frac{|x_i(t+\Delta t)|^2_{\mathcal
{L}(y(t+\Delta t))} -|x_i(t+\Delta t)|^2_{\mathcal
{L}(y(t))}}{\Delta t}\nonumber\\&&+\lim_{\Delta t\rightarrow
0}\frac{|x_i(t+\Delta t)|^2_{\mathcal
{L}(y(t))}-|x_i(t)|^2_{\mathcal {L}(y(t))}}{\Delta t}.
\end{eqnarray}
Then, according to (\ref{r10}), we obtain
\begin{eqnarray}\label{104}
&&\lim_{\Delta t\rightarrow 0}\frac{|x_i(t+\Delta t)|^2_{\mathcal
{L}(y(t))}-|x_i(t)|^2_{\mathcal {L}(y(t))}}{\Delta t}\nonumber\\
&=&\left.\frac{d }{ds}|x_i(s)|^2_{\mathcal {L}(y(t))} \right|_{s=t}\nonumber\\
&=& \left.\langle \nabla |x_i(s)|_{\mathcal {L}(y(t))}^2,\dot{x_i}(s)\rangle\right|_{s=t} \nonumber\\
&=&2\langle x_i(t)-\mathcal {P}_{\mathcal {L}(y(t))}(x_i(t)),
\sum_{j \in N_i(\sigma(t))}a_{ij}(x_j(t)-x_i(t))+\sum_{j \in
L_i(\sigma(t))}b_{ij}(y_j(t)-x_i(t))+w_i(t)\rangle.\nonumber\\
\end{eqnarray}
Furthermore, according to Lemma \ref{lem7},
$$
\lim_{\Delta t\rightarrow 0} \frac{|x_i(t+\Delta t)|_{\mathcal
{L}(y(t+\Delta t))} -|x_i(t+\Delta t)|_{\mathcal
{L}(y(t))}|}{\Delta t} \leq \lim_{\Delta t\rightarrow 0}
\frac{\int_{t}^{t+\Delta t}r(s)ds}{\Delta t}=r(t),
$$
and then it is easy to find that
\begin{eqnarray}\label{33}
\lim_{\Delta t\rightarrow 0}\frac{|x_i(t+\Delta t)|^2_{\mathcal
{L}(y(t+\Delta t))} -|x_i(t+\Delta t)|^2_{\mathcal
{L}(y(t))}}{\Delta t}&=& \lim_{\Delta t\rightarrow 0} \frac{|x_i(t+\Delta t)|_{\mathcal
{L}(y(t+\Delta t))} -|x_i(t+\Delta t)|_{\mathcal
{L}(y(t))}}{\Delta t}\nonumber\\
&&\ \ \cdot (|x_i(t+\Delta t)|_{\mathcal
{L}(y(t+\Delta t))} +|x_i(t+\Delta t)|_{\mathcal
{L}(y(t))})\nonumber\\
&\leq&2r(t) |x_i(t)|_{\mathcal {L}(y(t))}.
\end{eqnarray}
Therefore,
\begin{eqnarray}\label{e12}
\frac{d}{dt}\psi_i(t)
&\leq&2\langle x_i-\mathcal {P}_{\mathcal {L}(y(t))}(x_i),
\sum\limits_{j \in N_i(\sigma(t))}a_{ij}(x)(x_j-x_i)+\sum_{j \in
L_i(\sigma(t))}b_{ij}(x)(y_j-x_i)+w_i(t)\rangle\nonumber\\
&&+2r(t) |x_i(t)|_{\mathcal {L}(y(t))}.
\end{eqnarray}
Moreover, let $\mathcal{I}(t)$ denote the set containing all the agents
that reach the maximal distance away from $\mathcal {L}(y(t))$ at
time $t$. Then, for any $i \in \mathcal{I}(t)$, according to (\ref{r9}), one has
\begin{eqnarray}\label{31}
\langle x_i-\mathcal {P}_{\mathcal {L}(y(t))}(x_i),y_j-x_i\rangle&\leq&
\langle x_i-\mathcal {P}_{\mathcal {L}(y(t))}(x_i), y_j-\mathcal {P}_{\mathcal
{L}(y(t))}(x_i)\rangle\nonumber\\ &&+\langle x_i-\mathcal {P}_{\mathcal
{L}(y(t))}(x_i),
\mathcal {P}_{\mathcal {L}(y(t))}(x_i)-x_i\rangle\nonumber\\
&\leq& \langle x_i-\mathcal {P}_{\mathcal {L}(y(t))}(x_i),\mathcal {P}_{\mathcal {L}(y(t))}(x_i)-x_i\rangle\nonumber\\
&=& -\psi_i(t)
\end{eqnarray}
for any $j\in L_i({\sigma(t)})$. Furthermore, in light of Lemma
\ref{lem6}, since $i \in \mathcal{I}(t)$,
$$
\langle x_i-\mathcal {P}_{\mathcal {L}(y(t))} (x_i),x_j-x_i\rangle\leq
-|x_i|_{\mathcal {L}(y(t))}(|x_i|_{\mathcal
{L}(y(t))}-|x_j|_{\mathcal {L}(y(t))})\leq 0
$$
for any $j\in N_i({\sigma(t)})$. Therefore, the conclusion follows
since
\begin{eqnarray}
D^+\Psi(t)&=&\max_{i \in \mathcal{I}(t)} \frac{d}{dt}
\psi_i(t)\nonumber\\
&\leq&2\max_{i \in \mathcal{I}(t)} [\langle x_i-\mathcal {P}_{\mathcal {L}(y(t))}(x_i),
w_i(t)\rangle+2r(t) |x_i(t)|_{\mathcal {L}(y(t))}]\nonumber\\
&\leq& 2(r(t)+\max_{i\in\mathcal {V}_F}|w_i(t)|)\max_{i \in
\mathcal{I}(t)} |x_i(t)|_{\mathcal {L}(y(t))}\nonumber\\
&=& 2q(t)\sqrt{\Psi(t)}\nonumber
\end{eqnarray}
according to Lemma \ref{lem3}. \hfill$\square$

\section{Connectivity and SISS}

In this section, we study the SISS with respect to the convex set
spanned by the moving leaders in an important connectivity case,
uniformly jointly L-connected (UJLC) topology. Without loss of
generality, we will assume $n\geq 2$ in the sequel.

\subsection{Main results}

Suppose $z=(u^T,w^T)^T\in L_\infty$ in this section.  Then we have
the main result on SISS.

\begin{theorem}\label{thm5}
System (\ref{5}) is SISS  with respect to ${\cal L}(y(t))$ and with
$z$ as the input if and only if $\mathcal {G}_{\sigma(t)}$ is UJLC.
\end{theorem}

The main difficulties to obtain the SISS inequalities in the UJLC
case are how to estimate the convergence rate in a time interval by
``pasting" time subintervals together and how to estimate
the impact of the input $z$ to the agent motion.

To prove Theorem \ref{thm5}, we first present two lemmas to estimate
the distance error in the two standard cases during
$t\in[t_0,t_0+T_\ast]$ for $t_0\geq 0$ and a constant
$T_\ast>\tau_D$ with $\tau_D$ as the dwell time of switching.

\begin{lem}
\label{lem8} If there is an arc $(j,i)$ leaving from follower
$j\in\mathcal {V}_L$ entering $i\in\mathcal {V}_F$ in $\mathcal
{G}_{\sigma(t)}$ for all $t\in[t_0,t_0+\tau_D)$, then there exist a
continuous function $\mu(s):[0,T_\ast]\mapsto(0,1]$ and a constant
$\gamma_1>0$ such that
\begin{equation}
|x_i(t)|_{\mathcal {L}(y(t))}\leq \mu(t-t_0)|x(t_0)|_{\mathcal
{L}(y(t_0))} +\gamma_1\|z\|_{\infty},\;\forall
t\in[t_0,t_0+T_\ast].
\end{equation}
\end{lem}
Proof: See Appendix A.1. \hfill$\square$

\begin{lem}
\label{lem9} If there is an arc $(i,m)$ leaving from $i\in \mathcal
{V}_F$ entering $m\in \mathcal {V}_F$ in $\mathcal {G}_{\sigma(t)}$
for all $t\in[t_0,t_0+\tau_D)$, and
\begin{equation}\label{16}
|x_i(t)|_{\mathcal {L}(y(t))}\leq \mu_0 |x(t_0)|_{\mathcal
{L}(y(t_0))}+d_0,\;\forall t\in [t_0,t_0+\tau_D)
\end{equation}
for constants $\mu_0\in (0,1)$ and $d_0>0$, then there exist a
continuous function $\xi_{\mu_0}(s):[0,T_\ast]\mapsto(0,1]$ and a
positive constant $\gamma_2$ such that
\begin{equation}
\|x_m(t)\|_{\mathcal {L}(y(t))}\leq
\xi_{\mu_0}(t-t_0)|x(t_0)|_{\mathcal
{L}(y(t_0))}+\gamma_2\|z\|_{\infty}+d_0,\;\forall
t\in[t_0,t_0+T_\ast]
\end{equation}
\end{lem}
Proof: See Appendix A.2. \hfill$\square$

\begin{figure}[ht]
\centerline{\epsfig{figure=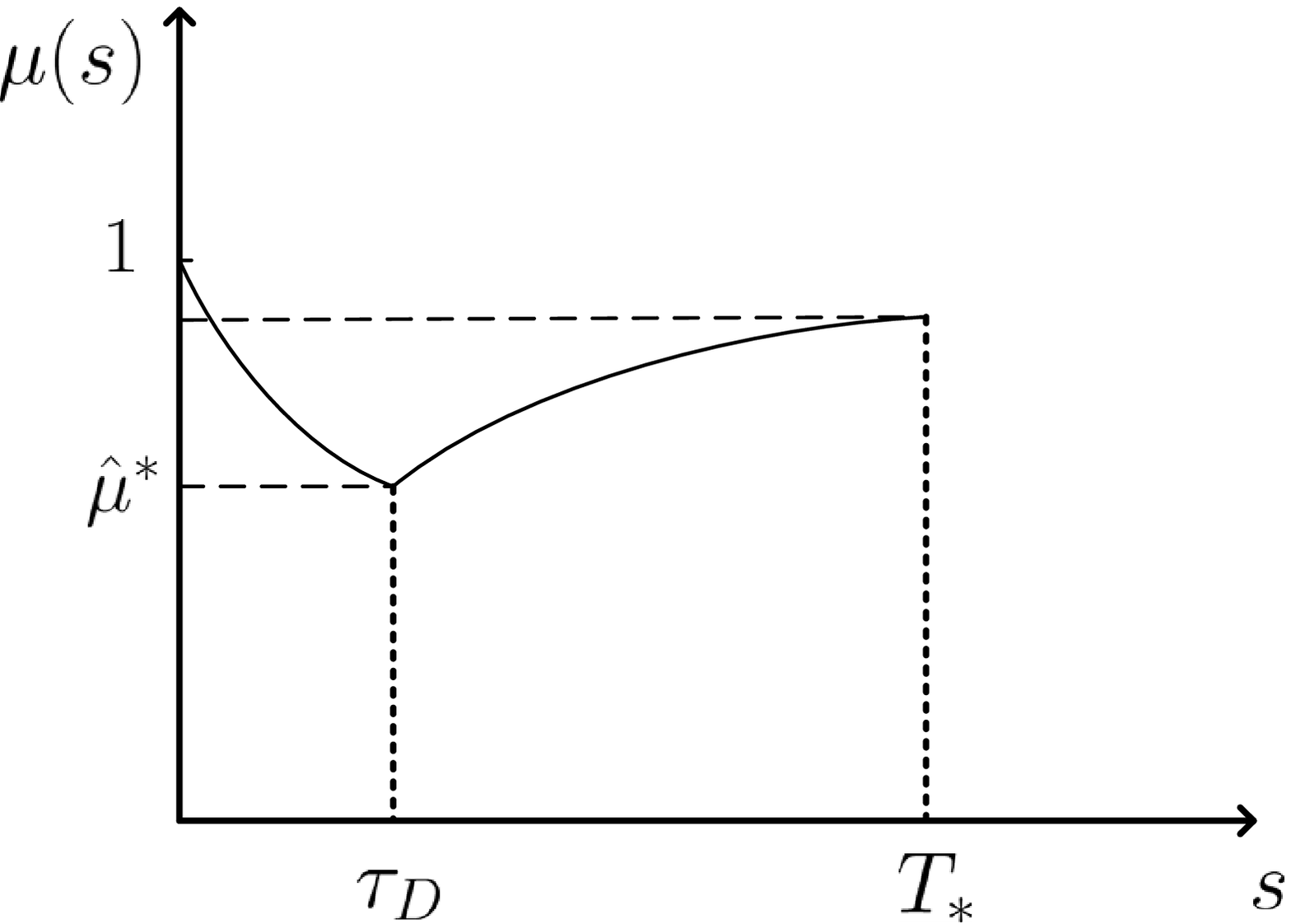, width=0.40\linewidth=0.2,
height=0.30\linewidth=0.2}\epsfig{figure=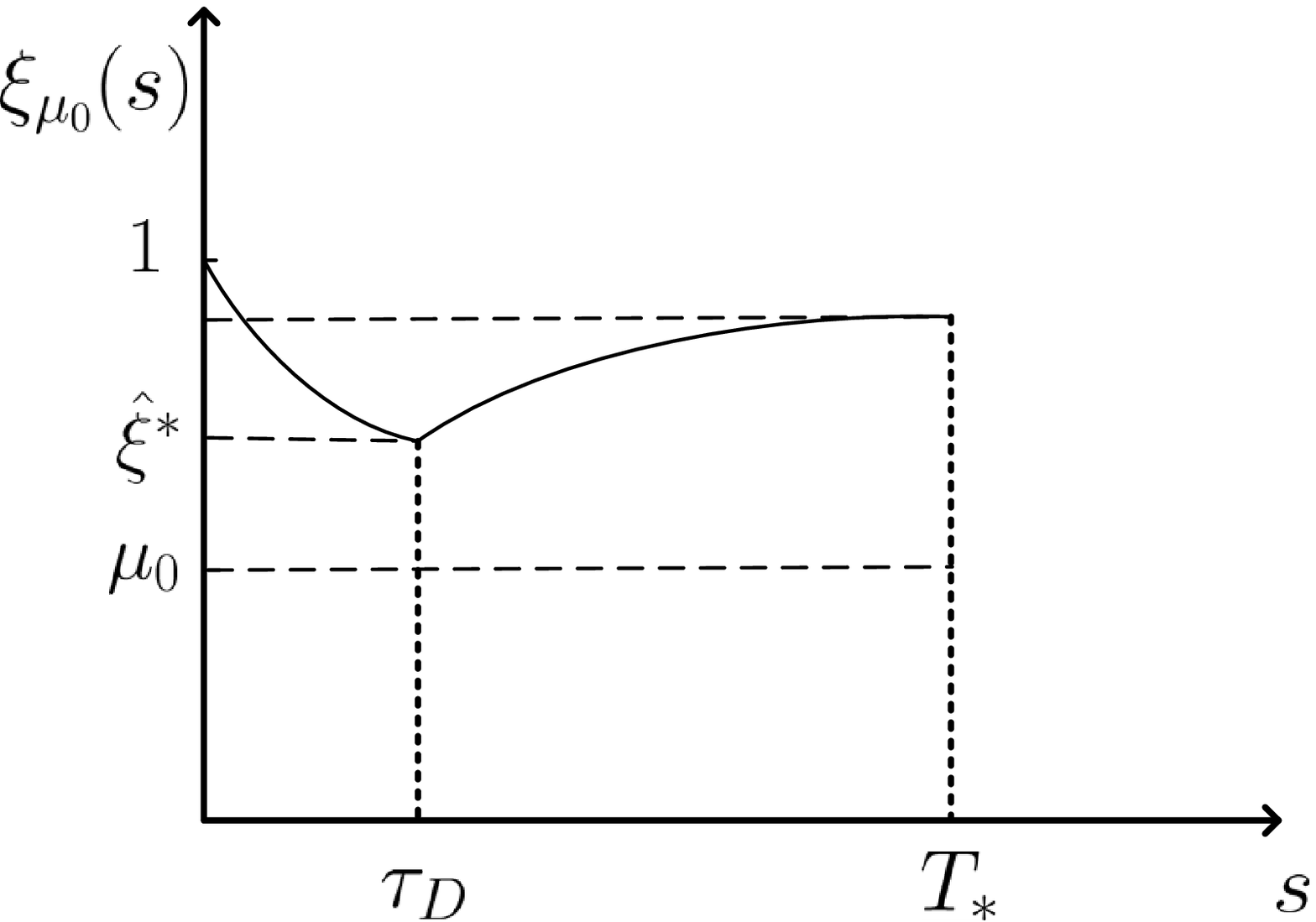,
width=0.40\linewidth=0.2,
height=0.30\linewidth=0.2}}\caption{$\mu(s)$ and
$\xi_{\mu_0}(s)$}\label{fig1}
\end{figure}

\begin{remark}
The following properties of $\mu(s)$ and $\xi_{\mu_0}(s)$ are quite
critical in the study of the set tracking with jointly L-connected
topology (see Fig. \ref{fig1}):
\begin{itemize}
\item[(i)] $\mu(0)=\xi_{\mu_0}(0)=1$.
\item [(ii)] $\mu(s)$ and $\xi_{\mu_0}(s)$ are strictly decreasing during $s\in[0,\tau_D]$.
\item [(iii)] $\mu(s)$ and $\xi_{\mu_0}(s)$ are strictly increasing during $s\in [\tau_D,T^\ast]$,
and $\mu(T^\ast)<1,\xi_{\mu_0}(T^\ast)<1$.
\end{itemize}
\end{remark}

%\begin{figure}[ht]
%\centerline{\epsfig{figure=figure2.eps, width=0.5\linewidth=0.25,
%height=0.3\linewidth=0.2}}\caption{The property of }\label{fig2}
%\end{figure}

Next, we introduce the following lemma to state an important property for UJLC graphs.

\begin{lem}
\label{lem5} If $\mathcal {G}_{\sigma(t)}$ is UJLC, then, for any
$t>0$ and $i\in \mathcal {V}_F$, there is a path $\widehat{(j,i)}$
from some leader $j\in\mathcal {V}_L$ to follower $i$ in $\mathcal
{G}([t,t+T_0))$ with $T_0\triangleq T+2\tau_D$, and each arc of
$\widehat{(j,i)}$ exists in a time interval with length $\tau_D$ at
least during $[t,t+T_0)$.
\end{lem}

\noindent Proof: Denote  $t_1$ as the first moment when the
interaction topology switches within $[t,t+T_0)$ (suppose there are
switchings without loss of generality). If $t_1\geq t+\tau_D$, then,
for any $i\in\mathcal {V}_F$, there is a path $\widehat{(j,i)}$ from
some leader with index $j\in\mathcal {V}_L$ to agent $i$ in
$\mathcal {G}([t,t+T))$, where each arc stays there for at least the
dwell time $\tau_D$ during $[t,t+T+\tau_D)$ due to the definition of
$\tau_D$. On the other hand, if $t_1<t+\tau_D$,
$t_1+T+\tau_D<t+T_0$. Then, for any $i\in\mathcal {V}_F$, there is
also a  path $\widehat{(j,i)}$ from some leader $j\in\mathcal {V}_L$
to agent $i$ in $\mathcal {G}([t_1,t_1+T))$ in $[t_1,t_1+T+\tau_D)$
with each arc exists for at least $\tau_D$. This completes the
proof. \hfill$\square$

\begin{remark}
If there is a convex set $\Omega$ such that ${\cal L}(y(t))\in
\Omega, \forall t\geq 0$, that is, $\Omega$ is a positively
invariant set for the leaders, then $|x(t)|_{\Omega}\leq
|x(t)|_{{\cal L}(y(t))}$.  By Theorem \ref{thm5}, system (\ref{5})
is SISS with respect to $\Omega$ with $w$ as the input if $\mathcal
{G}_{\sigma(t)}$ is UJLC. %Moreover, the uniform joint
%L-connectedness for $\mathcal {G}_{\sigma(t)}$ is also necessary for
%the SISS with respect to $\Omega$.
\end{remark}

Sometimes, the velocities of the moving leaders and uncertainties in
agent dynamics (maybe because of the online estimation) may vanish.
To be strict, consider the following condition
\begin{equation}
\label{C2}\begin{cases} \lim_{t\rightarrow
+\infty}u_i(y,t)=0\,\mbox{uniformly for}\ y;\; i=1,\dots, k;\\
\lim_{t\rightarrow +\infty}w_j(t)=0,\; j=1,\dots, n.\end{cases}
\end{equation}

Clearly, (\ref{C2}) yields that for any $\varepsilon>0$, there is
$T_\varepsilon>0$ such that $\|z^{T_\varepsilon}\|_\infty<
\varepsilon$, where $z^{T_\varepsilon}$ is the truncated part of $z$
defined on $[T_\varepsilon,+\infty)$. Suppose (\ref{C2}) holds and
$\mathcal {G}_{\sigma(t)}$ is UJLC. Based on Theorem \ref{thm5}, for
any $\varepsilon>0$, there is $T_\varepsilon>0$ such that
$$
|x(t)|_{\mathcal {L}(y(t))}\leq \beta(|
x(T_\varepsilon)|_{\mathcal {L}(y(T_\varepsilon))},t )+\gamma
(\varepsilon).
$$
Hence, the set tracking for system (\ref{5}) with respect to set
${\cal L}(y(t))$ is achieved easily. On the other hand, similar to
the proof of Theorem \ref{thm5}, the necessity of the global set
tracking for system (\ref{5}) with condition (\ref{C2}) can also be
simply proved by counterexamples since $|z(t)|$ may be large and the
distance error may accumulate to a very large value over a
sufficiently long period of time. Therefore, we have the following
result.

\begin{coro}\label{cor2}
The global set tracking with respect
to ${\cal L}(y(t))$ is achieved for all $z(t)$ satisfying (\ref{C2}) if and only if $\mathcal
{G}_{\sigma(t)}$ is UJLC.
\end{coro}

\subsection{Proof of Theorem \ref{thm5}}

We are now in a position to prove Theorem \ref{thm5}: ``If" part:
Denote $T_\ast=nT_0$ with $T_0=T+2\tau_D$. Then we
estimate $\Psi(t)$ at subintervals $[t^\ast+(j-1)T_0,t^\ast+jT_0]$
for $j=1,\dots,n$.

Based on Lemma \ref{lem5}, in $[t^\ast,t^\ast+T_0)$, there must be
an arc $({j_1},{i_1})\in \mathcal {E}([t^\ast,t^\ast+T_0))$ leaving
from a leader ${j_1}\in \mathcal {V}_L$ to a follower ${i_1}\in
\mathcal {V}_L$ and this arc remains for at least $\tau_D$. Suppose
$(j_1,i_1)\in \mathcal {E}_{\sigma(t)}$ for
$t\in[t_1,t_1+\tau_D)\subset [t^\ast,t^\ast+T_0)$. According to
Lemma \ref{lem8},
\begin{equation}
|x_{i_1}(t)|_{\mathcal {L}(y(t))}\leq \mu(t-t_1)|x(t_1)|_{\mathcal {L}(y(t_1))}
+\gamma_1\|z\|_{\infty},\; t\in [t_1,t_1+T_\ast],
\end{equation}
where $\mu(s)$ and $\gamma_1$ were defined in Lemma \ref{lem8}. Take
$\eta_1=\sup\limits\{\mu(s)\mid s\in [T_0,T_\ast]\}=\mu (T_\ast)$.
Since $0<\mu_1<1$,
\begin{equation}\label{20}
|x_{i_1}(t)|_{\mathcal {L}(y(t))}\leq \eta_1|x(t_1)|_{\mathcal {L}(y(t_1))}
+\gamma_1\|z\|_{\infty},\;t\in [t^\ast+T_0,t^\ast+T_\ast].
\end{equation}
Furthermore, in $[t^\ast+T_0,t^\ast+2T_0)$, there must be a follower
${i_2}\in\mathcal {V}_F,{i_2}\neq i_1$, such that there exists an
arc $({j_2},{i_2})$ for some ${j_2}\in\mathcal {V}_L$, or an arc
$({i_1},{i_2})$  in $\mathcal {E}([t^\ast+T_0,t^\ast+2T_0))$.

There are two cases:
\begin{itemize}
\item[1)] If $({j_2},{i_2})\in \mathcal {E}_{\sigma(t)}$ for
$t\in[t_2,t_2+\tau_D)\subset[t^\ast+T_0,t^\ast+2T_0)$, one also has
\begin{equation}
|x_{i_2}(t)|_{\mathcal {L}(y(t))}\leq \eta_1|x(t_2)|_{\mathcal {L}(y(t_2))}
+\gamma_1\|z\|_{\infty},\;t\in
[t^\ast+2T_0,t^\ast+T_\ast].
\end{equation}

\item[2)] If $({i_1},{i_2})\in \mathcal {E}_{\sigma(t)}$ for
$t\in[t_2,t_2+\tau_D)\subset[t^\ast+T_0,t^\ast+2T_0)$. According to (\ref{g2}) and
Lemma \ref{lem10}, one has
$$
|x(t_1)|_{\mathcal {L}(y(t_1))}\leq |x(t_2)|_{\mathcal {L}(y(t_2))}+ \sqrt{2}\|z\|_{\infty}\cdot|t_2-t_1|\leq|x(t_2)|_{\mathcal {L}(y(t_2))}+ 2\sqrt{2}\|z\|_{\infty} T_0,
$$
Thus, (\ref{20}) will lead to
\begin{equation}
|x_{i_1}(t)|_{\mathcal {L}(y(t))}\leq \eta_1|x(t_2)|_{\mathcal {L}
(y(t_2))}+(2\sqrt{2}\eta_1T_0+\gamma_1)\|z\|_{\infty},\;t\in [t^\ast+T_0,t^\ast+T_\ast].
\end{equation}

Then, by Lemma \ref{lem9}, if we take
$\eta_2=\xi_{\eta_1}((n-1)T_0)$, then
\begin{equation}
|x_{i_2}(t)|_{\mathcal {L}(y(t))}\leq \eta_2|x(t_2)|_{\mathcal
{L}(y(t_2))}+
(\gamma_2+2\sqrt{2}\eta_1T_0+\gamma_1)\|z\|_{\infty},\;
t\in [t^\ast+2T_0,t^\ast+T_\ast].
\end{equation}
\end{itemize}
Because $\eta_2>\eta_1$,
\begin{equation}\label{19}
|x_{\jmath}(t)|_{\mathcal {L}(y(t))}\leq \eta_2|x(t_2)|_{\mathcal
{L}(y(t_2))}+(\gamma_2+2\sqrt{2}\eta_1T_0+\gamma_1)\|z\|_{\infty},\;\jmath=i_1,i_2,
t\in [t^\ast+2T_0,t^\ast+T_\ast].
\end{equation}

Repeating the above procedure yields
$$
\eta_j=\xi_{\eta_{j-1}}((n-j+1)T_0),\;j=3,\dots,n
$$
and $t_j\in [t^\ast+jT_0,t^\ast+T_\ast)$ such that, there exists ${i_j}\in\mathcal {V}_F,\;j=3,\dots,n$ satisfying
\begin{equation}\label{18}
|x_{\jmath}(t)|_{\mathcal {L}(y(t))}\leq
\eta_j|x(t_j)|_{\mathcal
{L}(y(t_j))}+[(j-1)\gamma_2+2\sqrt{2}\sum_{l=1}^{j-1}\eta_l
T_0+\gamma_1]\|z\|_{\infty},\;\jmath=i_1,\dots,i_j
\end{equation}
for $t\in [t^\ast+jT_0,t^\ast+T_\ast]$.
Moreover, the nodes $i_j,j=1,2,\dots,n$ are distinct.

Denote $\eta_\ast =\eta_n$, and then $0<\eta_\ast<1$. Thus,
(\ref{18}) leads to
\begin{equation}
|x_{\jmath}(t^\ast+T_\ast)|_{\mathcal {L}(y(t^\ast+T_\ast))}\leq
\eta_\ast|x(t^\ast)|_{\mathcal {L}(y(t^\ast))}+[(1+2\sqrt{2})\eta_\ast
T_\ast+(n-1)\gamma_2+\gamma_1]\|z\|_{\infty},
\end{equation}
for any $\jmath\in\mathcal {V}_F$, which leads to
\begin{equation}\label{22}
|x(t^\ast+T_\ast)|_{\mathcal {L}(y(t^\ast+T_\ast))}\leq
\eta_\ast|x(t^\ast)|_{\mathcal {L}(y(t^\ast))}+[(1+2\sqrt{2})\eta_\ast
T_\ast+(n-1)\gamma_2+\gamma_1]\|z\|_{\infty}.
\end{equation}
Therefore, $\forall N=1,2,\dots$,
\begin{equation}\label{26}
|x(N T_\ast)|_{\mathcal {L}(y(N T_\ast))}\leq
\eta_\ast^N|x^0|_{\mathcal {L}(y^0)}+\sum_{j=0}^{N-1}\eta_\ast^j[(1+2\sqrt{2})\eta_\ast
T_\ast+(n-1)\gamma_2+\gamma_1]\|z\|_{\infty}.
\end{equation}
Again by Lemma \ref{lem10}, one has
\begin{equation}
|x(t)|_{\mathcal {L}(y(t))}\leq \beta(|x^0|_{\mathcal
{L}(y^0)},t)+\gamma(\|z\|_{\infty})
\end{equation}
with
$$
\beta(|x^0|_{\mathcal {L}(y^0)},t)\triangleq\eta_\ast^{\lfloor
\frac{t}{T^\ast}\rfloor}|x^0|_{\mathcal {L}(y^0)},\quad
\gamma(s)\triangleq[\frac{(1+2\sqrt{2})\eta_\ast T_\ast+(n-1)\gamma_2+\gamma_1
}{1-\eta_\ast}+T_\ast] s
$$
where $\lfloor \frac{t}{T^\ast}\rfloor$ denotes the largest integer
no greater than $\frac{t}{T^\ast}$, which implies the conclusion.

``Only if" part: If $\mathcal {G}_{\sigma(t)}$ is not UJLC, there is
a time sequence $ 0<T_1<T_2<\dots $ such that $\mathcal
{G}([T_{2\kappa-1},T_{2\kappa}))$ is not L-connected for
$\kappa=1,2,\dots$ with $\lim_{\kappa\rightarrow \infty}
(T_{2\kappa}-T_{2\kappa-1})=\infty$.  Taking
$x_i(0)=(0,\dots,0)^T\in R^d,\forall i\in \mathcal {V}_F$ and
$y_i(0)=(1,\dots,1)^T\in R^d, \forall i\in \mathcal {V}_L$ with
$w_i(t)\equiv 0,\forall i\in \mathcal {V}_F$ and $u_i(y,t)\equiv
(1,\dots,1)^T,\forall i\in \mathcal {V}_L$, we obtain $\mathcal
{L}(y(t))=\{(1+t,\dots,1+t)^T\}$. Since $\mathcal
{G}([T_{2\kappa-1},T_{2\kappa}))$ is not L-connected, there is $i\in
\mathcal {V}_F$ such that agent $i$ is reachable from no leader.
Define $\hat{\mathcal {V}}_i^1\triangleq \{j\in\mathcal {V}|i\; \mbox{is
reachable from}\; j\ \mbox{in graph}\ \mathcal {G}([T_{2\kappa-1},T_{2\kappa}))\}$.
Since $\hat{\mathcal {V}}_i^1$ contains no leader and there is no arc
entering $\hat{\mathcal {V}}_i^1$, no agent in $\hat{\mathcal {V}}_i^1$ leaves
$co\{x_\jmath(T_{2\kappa-1}),\jmath\in\hat{\mathcal {V}}_i^1\}$ when
$t\in[T_{2\kappa-1},T_{2\kappa})$. Moreover, none of the followers
can enter $\mathcal {L}(y(t))$ in finite time. Therefore,
$$
\lim_{\kappa\to \infty}|x_\jmath(T_{2\kappa})|_{\mathcal
{L}(y(T_{2\kappa}))}\geq \lim_{\kappa\to \infty}
(T_{2\kappa}-T_{2\kappa-1})=+\infty,\quad \forall \jmath\in \hat{\mathcal {V}}_i^1.
$$
Thus, the SISS with respect to $\mathcal {L}(y(t))$ cannot be
achieved. \hfill$\square$

%\begin{remark}
%The distance error can be estimated once the upper bound of the
%uncertain inputs is known; namely, if
%$\|z\|_{\infty}\leq \bar c$ and $\mathcal
%{G}_{\sigma(t)}$ is uniformly jointly L-connected, then
%$$
%\limsup_{t\to \infty} \|x(t)\|_{\mathcal {L}(y(t))}\leq \tilde c=[
%\frac{2\eta_\ast T_\ast  +(n-1)\gamma_2 +\gamma_1}{1-\eta_\ast}+T_\ast]\bar
%c.
%$$
%As $a_*, a^*, b_*$ tend to infinity, we cannot ensure $\tilde c\to
%0$.  In fact, $ \gamma_1\to T_\ast$ and $\gamma_2 \to
%\frac{n-1}{n-2}T_\ast$ as $a^*\to \infty$, and the limit of
%$\eta_\ast$ is some value between 0 and 1, though its explicit
%expression is complicated, even depending on how fast $a_*, a^*,
%b_*$ tend to infinity. This observation shows that the set tracking
%is hardly to be achieved no matter how large the weights of the
%information flows are in general cases of uniformly jointly
%L-connected interaction topologies.
%\end{remark}

\section{Connectivity and SiISS}

In this section, we aim at the connectivity requirement to ensure
the set integral-input-to-state stability (SiISS) when $\mathcal
{G}_{\sigma(t)}$ is jointly L-connected (JLC).

\subsection{Main results}
Theorem \ref{thm5} showed an equivalent relationship between SISS
and UJLC. However, this is not true for SiISS.  Here, we propose a
couple of theorems about SiISS. The proofs of these conclusions can
be found in the following subsection.

First of all, we propose a sufficient condition.

\begin{theorem}\label{thmr1}
\label{thm8} System (\ref{5}) is SiISS with respect to $\mathcal
{L}(y(t))$ if $\mathcal {G}_{\sigma(t)}$ is UJLC.
\end{theorem}

\begin{remark}
\label{rem3}  JLC of $\mathcal {G}_{\sigma(t)}$ (i.e., $\mathcal
{G}([t,\infty))$ is L-connected for any $t$) is necessary for the
SiISS, though it is not sufficient. If $\mathcal
{G}([\tilde{T},\infty))$ is not L-connected for some $\tilde{T}>0$,
there is a subset $\hat{\mathcal {V}}_F\subseteq \mathcal {V}_F$
such that no arcs enter $\hat{\mathcal {V}}_F$ in $\mathcal
{G}([\tilde{T},\infty))$. Hence, the agents in $\hat{\mathcal
{V}}_F$ may not be SiISS for some initial conditions since they will
not be influenced by the convex leader-set after $\tilde{T}$.
\end{remark}

UJLC, which is a special case of JLC, provides a sufficient
condition for SiISS, but UJLC is not necessary to ensure SiISS. In
fact, there are other cases of JLC to make SiISS hold. Here we
consider two important special JLC cases i.e., bidirectional
graphs and acyclic graphs.

A digraph $\mathcal {G}$ is called a bidirectional graph
when $i$ is a neighbor of $j$ if and only if $j$ is a neighbor of
$i$, but the weight of arc $(i,j)$ may not be equal to that of arc
$(j,i)$.   The next result shows a necessary and
sufficient condition for the bidirectional case.

\begin{theorem}\label{thm10}
Suppose that $\mathcal {G}^F_{\sigma(t)}$ is bidirectional for all $t\geq 0$. Then
system (\ref{5}) is SiISS if and only if $\mathcal {G}_{\sigma(t)}$
is JLC.
\end{theorem}

The next lemma shows an important property for an acyclic digraph,
that is, a digraph without cycles.

\begin{lem}
\label{claim}  Assume that $\mathcal
{G}^F([0,+\infty))$ is acyclic and $\mathcal {G}([0,+\infty))$ is L-connected. Then there is a partition of $\mathcal {V}_F$ by
$\mathcal {V}_F=\bigcup_{i=1}^{k_0} \mathcal {V}_i^F, k_0\geq 1$ such
that in graph $\mathcal {G}([0,+\infty))$,  all the arcs  entering node set
$\mathcal {V}_1^F$ are from $\mathcal {V}_L$ ; and all the arcs entering node set $\mathcal {V}_j^F,j=2,\dots,k_0$ are
from $\mathcal {V}_L \cup (\bigcup_{i=1}^{j-1} \mathcal
{V}_i^F)$.
\end{lem}

\noindent Proof: First we prove $\mathcal {V}^F_1$ exists by contradiction. If $\mathcal {V}^F_1$ does not exist, every agent
$i,i\in \mathcal {V}_F$ has neighbors within $\mathcal {V}_F$ in
$\mathcal {G}([0,+\infty))$. Denote $\hat{\mathcal {V}}^F_1\triangleq
\{j\in\mathcal {V}_F |\mbox{there\ is\ an\ arc\ leaving\ from}\ \mathcal {V}_L\ \mbox{entering}\ j \}$. Clearly $\hat{\mathcal {V}}^F_1\neq
\emptyset$. Take $i_0\in \hat{\mathcal {V}}^F_1$. Then, there is
$j_1\in\mathcal {V}_F$ such that $(j_1,i_0)\in \mathcal
{G}([0,+\infty))$. Moreover, we can associate $j_1$ with
$i_1\in\hat{\mathcal {V}}^F_1 $ ($i_1$ cannot be $i_0$, of course)
such that there is a path $\widehat{(i_1,j_1)}$ in $\mathcal
{G}([0,\infty))$ ($i_1=j_1$ if $j_1\in\hat{\mathcal {V}}^F_1$). Hence,
a path $\widehat{(i_1,i_0)}$ in $\mathcal {G}([0,+\infty))$ is
found. Regarding $i_1$ as $i_0$ and repeating the above procedure
yields the existence of $\widehat{(i_2,i_1)}$ in $\mathcal
{G}([0,+\infty))$ with $i_2 \in\hat{\mathcal {V}}^F_1 $. In this way,
we obtain a path $\widehat{(i_{l+1},i_l})$ in $\mathcal
{G}([0,+\infty))$ with $i_l \in\hat{\mathcal {V}}^F_1,l=2,3,\dots $.
Since the nodes in $\hat{\mathcal {V}}^F_1$ are finite, there has to
be $i_{l_1}=i_{l_2}$ for some $l_1>l_2\geq 0$, which lead to a
directed cycle in $\mathcal {G}([0,+\infty))$. Therefore, there is
$\mathcal {V}^F_1$ to make the conclusion hold.

Next, by replacing
$\mathcal {V}_L$ with $\mathcal {V}^F_1\cup\mathcal {V}_L$ in
$\mathcal {G}([0,\infty))$, with the same analysis we can find
$\mathcal {V}^F_2$ to make the conclusion hold.  Repeating this
procedure, since the number of all the agents is finite, there will
be a constant $k_0\geq 1$ such that $\mathcal {V}_F=\bigcup_{i=1}^{k_0} \mathcal {V}^F_i$. This completes the proof.
\hfill$\square$

Then we have a SiISS result for the acyclic graph case.

\begin{theorem}\label{thm12}
Assume that $\mathcal
{G}^F([0,+\infty))$ is acyclic. Then system
(\ref{5}) is SiISS if and only if $\mathcal {G}_{\sigma(t)}$ is JLC.
\end{theorem}

Furthermore, consider the following inequality
\begin{equation}
\label{C1} \int_0^{+\infty}|z(t)|dt<\infty.
\end{equation}

%\begin{remark}
%\label{rem-c1} If $(u,w)$ satisfies (\ref{C1}), then, for any time
%sequence $0=T_1<T_2 \dots < T_i < \dots$,
%\begin{equation}
%\label{pi} \sum_{i=1}^{\infty} p_i<\infty,\quad
%p_i=\int_{T_{i}}^{T_{i+1}}q(t)dt,\;i=1,2,\dots.
%\end{equation}
%\end{remark}

It is not hard to obtain the following results based on Theorems
\ref{thmr1}, \ref{thm10}, and \ref{thm12}. The proofs are omitted
for space limitations.

\begin{coro}\label{cor3}
System (\ref{5}) achieves the set tracking if (\ref{C1}) holds and
$\mathcal {G}_{\sigma(t)}$ is UJLC.
\end{coro}

\begin{coro}\label{cor4}
Suppose (\ref{C1}) holds with either $\mathcal {G}^F_{\sigma(t)}$ being bidirectional for all $t\geq 0$  or  $\mathcal
{G}^F([0,+\infty))$ being acyclic.
Then system (\ref{5}) achieves the global set tracking  if and only
if $\mathcal {G}_{\sigma(t)}$ is JLC.
\end{coro}
\begin{remark}
In general, the condition (\ref{C2}) does not imply and is not
implied by the condition (\ref{C1}). In fact, the considered leaders
converge to some points with (\ref{C1}), but the leaders can go to
infinity with (\ref{C2}). However, if $z(t)$ is uniformly continuous
in $[0,+\infty)$ (which can be guaranteed once $\dot{z}(t)$ is
bounded for $t\in [0,+\infty)$), (\ref{C2}) will then be implied by
(\ref{C1}) according to Barbalat's Lemma.
%and the uncertainties in the multi-agent system become vanishing.
%Therefore, the conclusions
%based on (\ref{C1}) can be directly applied to uncertainty-free
%multi-agent systems with stationary leaders.
\end{remark}

\begin{remark}
Corollaries \ref{cor3} and \ref{cor4} are consistent with
Proposition 6 in \cite{sontag1}, where (\ref{C1}) and integral-ISS
together resulted in the state stability. Moreover, the two
corollaries are also consistent with Theorems 15 and 17 in
\cite{shi09}, respectively, when $z\equiv 0$. However, different
from the limit-set-based approach given in \cite{shi09}, the
proposed method by virtue of (\ref{49}) and (\ref{120}) also
provides the estimation of the convergence rate.
\end{remark}

\begin{remark}
\label{rem4} Theorems \ref{thm5} and \ref{thmr1} with Remark
\ref{rem3} proved that for system (\ref{5}), SISS is equivalent to
UJLC, which implies SiISS, while JLC is a necessary condition,
namely,
$$
SISS\Longleftrightarrow UJLC \Longrightarrow SiISS \Longrightarrow
JLC.
$$
Thus, $ SISS \Longrightarrow SiISS, $ which is consistent with
Corollary 4 of \cite{sontag1}, where ISS implies iISS. Moreover,
Theorems \ref{thm10} and \ref{thm12} show that, in either
bidirectional or acyclic case,
$$
SiISS \Longleftrightarrow JLC.
$$
\end{remark}

\begin{remark}
As for set tracking (ST), Corollary \ref{cor2} shows that
$$
UJLC \Longleftrightarrow  ST, \forall \mbox{z(t) satifying}\ (\ref{C2}).
$$
Moreover, Corollaries \ref{cor3} and
\ref{cor4} show that as long as (\ref{C1}) holds,
$$
UJLC \Longrightarrow ST
$$
in general directed cases, and
$$
\quad JLC \Longleftrightarrow ST
$$
in either bidirectional or acyclic case.  Usually, SISS goes with
(\ref{C2}) and SiISS with (\ref{C1}), consistent with discussions on
ISS and iISS \cite{sontag1, sontag-wang}. Additionally, it is worth
pointing out that the differences between the statements in
Corollaries \ref{cor2} and those in \ref{cor3} result from the fact
that UJLC is necessary for SISS, but not necessary to SiISS.
\end{remark}
Although our results are consistent with the results on conventional
ISS or iISS, the analysis methods given in \cite{sontag1,
sontag-wang} are mainly based on an equivalent ISS-Lyapunov
function, which cannot be applied to our cases with a moving set and
switching topologies.

\subsection{Proofs}

To establish the SiISS in the JLC case, we will analyze the impact
of the integral of input $z(t)$ in a time interval and estimate the
convergence rates during this time interval by ``pasting" different
time subintervals together within the interval. The following lemmas
are given to estimate the convergence rates in different cases.

\begin{lem}\label{lem11}
If there is an arc $(j,i)$ leaving from $j\in\mathcal {V}_L$
entering $i\in\mathcal {V}_F$ in $\mathcal {G}_{\sigma(t)}$ for
$t\in[t_0,t_0+\tau_D)$, then there exists a strictly decreasing
function $\delta(s):[0,\tau_D]\mapsto(0,1]$ with $\delta(0)=1$  such
that
\begin{equation}
|x_i(t)|_{\mathcal {L}(y(t))}\leq
\delta(t-t_0)|x(t_0)|_{\mathcal {L}(y(t_0))}
+2\sqrt{2}\int_{t_0}^{t_0+\tau_D}|z(s)|ds,\;  t\in[t_0,t_0+\tau_D].
\end{equation}
\end{lem}

\noindent Proof: According to Lemma \ref{lem10}, $ \psi_j(t)\leq
\sqrt{\Psi(t)}\leq \sqrt{\Psi(t_0)}
+\int_{t_0}^{t}\sqrt{2}|z(s)|ds,\; j=1,...,n$ for any $t>t_0>0$.
Since there is an arc $(j,i)$ with $j\in\mathcal {V}_L,i\in\mathcal
{V}_F$ in $\mathcal {G}_{\sigma(t)}$ for $t\in[t_0,t_0+\tau_D)$,
$$
\frac{d}{dt}\psi_i(t) \leq -2b_\ast
\psi_i(t)+2\sqrt{2}|z(t)|\sqrt{\psi_i(t)}+2\langle x_i-\mathcal {P}_{\mathcal
{L}(y(t))}(x_i), \sum\limits_{j \in
N_i(\sigma(t))}a_{ij}(x)(x_j-x_i)\rangle.
$$
Based on Lemma \ref{lem6}, when $t\in [t_0,t_0+\tau_D)$,
\begin{eqnarray}
\langle x_i(t)-\mathcal {P}_{\mathcal {L}(y(t))}(x_i(t)),x_j(t)-x_i(t)\rangle
&\leq&  \sqrt{\psi_i(t)}
(\sqrt{\Psi(t)}-\sqrt{\psi_i(t)})\nonumber\\
 &\leq& \sqrt{\psi_i(t)}(\sqrt{\Psi(t_0)}+\int_{t_0}^{t}\sqrt{2}|z(s)|ds-\sqrt{\psi_i(t)} )\nonumber
\end{eqnarray}
Therefore,
$$
\frac{d}{dt}\psi_i(t)\leq
-2[b_\ast+(n-1)a^\ast]\psi_i(t)+2[\sqrt{2}|z(t)|
+(n-1)a^\ast(\sqrt{\Psi(t_0)}+\int_{t_0}^{t}\sqrt{2}|z(s)|ds)]\sqrt{\psi_i(t)},
$$
or equivalently,
$$
\frac{d}{dt}\sqrt{\psi_i(t)} \leq
-\lambda\sqrt{\psi_i(t)}+[\sqrt{2}|z(t)|+(n-1)a^\ast(\sqrt{\Psi(t_0)}+\int_{t_0}^{t_0+\tau_D}\sqrt{2}|z(s)|ds)]
$$
where $\lambda\triangleq b_\ast+(n-1)a^\ast$ for $t\in
[t_0,t_0+\tau_D)$.  Thus,
$$
\sqrt{\psi_i(t)} \leq \delta(t-t_0)\sqrt{\Psi(t_0)}
+\frac{b_\ast+2(n-1)a^\ast}{\lambda}\int_{t_0}^{t_0+\tau_D}\sqrt{2}|z(s)|ds,\;\;t\in
[t_0,t_0+\tau_D)
$$
with $\delta(s)\triangleq \frac{b_\ast e^{-\lambda
s}+(n-1)a^\ast}{\lambda},\, s\in[0,\tau_D]$, which implies the
conclusion. \hfill$\square$

\begin{lem}\label{lem12}
Suppose there is an edge $(i,m)$ leaving from $i\in \mathcal {V}_F$
entering $m\in \mathcal {V}_F$ in $\mathcal {G}_{\sigma(t)}$ and $
|x_i(t)|_{\mathcal {L}(y(t))}\leq \delta_0 |x(t_0)|_{\mathcal
{L}(y(t_0))}+\tilde c_0$ with constants $\delta_0\in (0,1)$ and
$\tilde c_0>0$ when $t\in [t_0,t_0+\tau_D)$.  Then there is a
strictly decreasing function
$\varphi_{\delta_0}(s):[0,\tau_D]\mapsto(0,1]$ with
$\varphi_{\delta_0}(0)=1$ such that
\begin{equation}
|x_m(t)|_{\mathcal {L}(y(t))}\leq \varphi_{\delta_0}
(t-t_0)|x(t_0)|_{\mathcal {L}(y(t_0))}+\tilde
c_0+2\sqrt{2}\int_{t_0}^{t_0+\tau_D}|z(s)|ds,\;
t\in[t_0,t_0+\tau_D].
\end{equation}
\end{lem}

\begin{lem}\label{lem13}
Given a constant $\hat{T}>0$, if there is $t_1\geq t_0$ with $
\|x_i(t_1)\|_{\mathcal {L}(y(t_1))}\leq \varepsilon_0
|x(t_0)|_{\mathcal {L}(y(t_0))}+\hat c_0$ for constants
$\varepsilon_0\in (0,1)$ and $\hat c_0>0$, then there is a  strictly
increasing function $\phi_{\varepsilon_0}(s):[0,\hat{T}]\mapsto
[\varepsilon_0,1)$  with $\phi_{\varepsilon_0}(0)=\varepsilon_0$
such that
\begin{equation}
|x_i(t)|_{\mathcal {L}(y(t))}\leq \phi_{\varepsilon_0}
(t-t_1)|x(t_0)|_{\mathcal {L}(y(t_0))}+\hat
c_0+2\sqrt{2}\int_{t_0}^{t_1+\hat{T}}|z(s)|ds,\;
t\in[t_1,t_1+\hat{T}],
\end{equation}
where $\phi_{\varepsilon_0}(s)=1-e^{-(n-1)a^\ast
s}(1-\varepsilon_0)$.
\end{lem}

The proofs of Lemmas \ref{lem12} and \ref{lem13} are similar to that
of Lemma \ref{lem11}, and therefore, omitted.

\begin{lem}\label{lem14}
Suppose $\mathcal {V}_F^1\subset \mathcal {V}_F$ is an nonempty
subset. If there are no arcs leaving from ${V}_F\setminus \mathcal
{V}_F^1$ entering $\mathcal {V}_F^1$ in $\mathcal
{G}([t_1,t_1+\hat{T}))$ for a given constant $\hat T>0$ and $
\|x_i(t_1)\|_{\mathcal {L}(y(t_1))}\leq \varepsilon_0
|x(t_0)|_{\mathcal {L}(y(t_0))}+\hat c_0,\;\forall i\in \mathcal
{V}_F^1 $ for constants $\varepsilon_0\in (0,1)$ and $\hat c_0>0$,
then
\begin{equation}
|x_i(t)|_{\mathcal {L}(y(t))}\leq \varepsilon_0
|x(t_0)|_{\mathcal {L}(y(t_0))}+\hat
c_0+\sqrt{2}\int_{t_1}^{t_1+t}|z(s)|ds,\quad
\end{equation}
\end{lem}

Taking $\Psi_1(t)=\max\limits_{i\in\mathcal {V}_F^1}\{ \psi_i(t)\}$
gives $ D^+\sqrt{\Psi_1(t)}\leq \sqrt{2}|z(t)|$ for
$t\in[t_1,t_1+\hat{T}]$ by virtue of the analysis given for Lemma
\ref{lem10}. Then Lemma \ref{lem14} can be obtained straightforwardly.

%{\noindent Proof:} When $t\in [t_1,\hat{T})$, one has
%\begin{eqnarray}
%\frac{d}{dt}\psi_i(t) &\leq& 2\sqrt{2}|z(t)|
%\sqrt{\psi_i(t)}+2\langle x_i-\mathcal {P}_{\mathcal {L}(y(t))}(x_i),
%\sum\limits_{j \in N_i(\sigma(t))}a_{ij}(x)(x_j-x_i)\rangle\nonumber\\
%&\leq& 2\sqrt{2}|z(t)| \sqrt{\psi_i(t)}
%+2(n-1)a^\ast\sqrt{\psi_i(t)}(\sqrt{\Psi(t_0)}-\sqrt{\psi_i(t)}
%+\int_{t_0}^{t_1+\hat{T}}\sqrt{2}|z(s)|ds)\nonumber
%\end{eqnarray}
%which is equivalent to
%$$
%\frac{d}{dt}\sqrt{\psi_i(t)}
%\leq-(n-1)a^\ast\sqrt{\psi_i(t)}+\sqrt{2}|z(t)|+(n-1)a^\ast(\sqrt{\Psi(t_0)}+\int_{t_0}^{t_1+\hat{T}}\sqrt{2}|z(t)|dt).
%$$
%Therefore,
%\begin{eqnarray}
%\sqrt{\psi_i(t)}
%&\leq& e^{-(n-1)a^\ast(t-t_1)}[\varepsilon_0 \sqrt{\Psi(t_0)}+\hat c_0]\nonumber\\
%&&+(1-e^{-(n-1)a^\ast(t-t_1)})[\sqrt{\Psi(t_0)}
%+\int_{t_0}^{t_1+\hat{T}}\sqrt{2}|z(t)|dt]+\int_{t_1}^{t}\sqrt{2}|z(s)|ds.\nonumber
%\end{eqnarray}
%Then the conclusion follows. \hfill$\square$

Now we are ready to prove Theorems \ref{thmr1}, \ref{thm10}, and
\ref{thm12}.

\noindent{\bf Proof of Theorem \ref{thmr1}}: Denote $T_\ast=nT_0$
with $T_0=T+2\tau_D$ defined in Lemma \ref{lem5}. If $\mathcal
{G}([t^\ast,t^\ast+T_0))$ is L-connected, there has to be an arc
$(j_1,i_1)\in \mathcal {E}_{\sigma(t)}$ for
$t\in[t_1,t_1+\tau_D)\subset [t^\ast,t^\ast+T_0)$ leaving from a
leader ${j_1}\in \mathcal {V}_L$ entering ${i_1}\in \mathcal {V}_L$
and this arc is kept there for a period of at least $\tau_D$.
Invoking Lemmas \ref{lem11} and \ref{lem13},
$$
|x_{i_1}(t)|_{\mathcal {L}(y(t))}\leq c_1|x(t_1)|_{\mathcal
{L}(y(t_1))} +4\sqrt{2}\int_{t^\ast}^{t^\ast+T_\ast}|z(s)|ds,\;\;
t\in [t_1,t^\ast+T_\ast],
$$
where $c_1=\phi_{\delta(\tau_D)}(T_\ast)$.

Furthermore, when $t\in [t^\ast+T_0,t^\ast+2T_0)$, there must be a
follower ${i_2}\in\mathcal {V}_F,{i_2}\neq i_1$ such that there
exists an arc $({j_2},{i_2})$ for some ${j_2}\in\mathcal {V}_L$, or
an arc $({i_1},{i_2})$ when $t\in[t_2,t_2+\tau_D)\subset
[t^\ast+T_0,t^\ast+2T_0))$. According to Lemmas \ref{lem12} and
\ref{lem13},
$$
|x_{i_2}(t)|_{\mathcal {L}(y(t))}\leq c_2|x(t_1)|_{\mathcal
{L}(y(t_1))} +8\sqrt{2}\int_{t^\ast}^{t^\ast+T_\ast}|z(s)|ds,\; t\in
[t_2,t^\ast+T_\ast],
$$
where $c_2=\phi_{\varphi_2}(T_\ast)$ with
$\varphi_2=\varphi_{c_1}(\tau_D)$.

Repeating the above procedure yields
$$
|x_{i_\ell}(t)|_{\mathcal {L}(y(t))}\leq
c_\ell|x(t_1)|_{\mathcal {L}(y(t_1))}
+4\sqrt{2}\ell\int_{t^\ast}^{t^\ast+T_*}|z(s)|ds,\; t\in [t^\ast+\ell T_0,t^\ast+T_\ast].
$$
for ${i_\ell}\in\mathcal {V}_F,\;\ell=3,\dots,n$, where
\begin{equation}\label{r1}
c_\ell=\phi_{\varphi_{\ell-1}}(T_\ast),\varphi_{\ell}=\varphi_{c_{\ell-1}}(\tau_D),
[t_\ell,t_\ell+\tau_D)\subset[t^\ast+(\ell-1)T_0,t^\ast+\ell T_0)
\end{equation} Moreover, the nodes of ${i_\ell},\ell=1,2,\dots, n$ are distinct.

Denote $\hat{c}\triangleq c_n$ from (\ref{r1}). Then we obtain
\begin{equation}
|x(t^\ast+T_\ast)|_{\mathcal {L}(y(t^\ast+T_\ast))}\leq
\hat{c}|x(t^\ast)|_{\mathcal
{L}(y(t^\ast))}+(4n+1)\sqrt{2}\int_{t^\ast}^{t^\ast+T_\ast}|z(s)|ds.
\end{equation}
It follows immediately that
\begin{equation}\label{49}
|x(K T_\ast)|_{\mathcal {L}(y(K T_\ast))}\leq \hat{c}^K
|x^0|_{\mathcal {L}(y^0)}+(4n+1)\sqrt{2}\sum_{j=1}^{K}\int_{(j-1)T_\ast}^{jT_\ast}\hat{c}^{K -j}|z(s)|ds
,\quad K=1,2,\dots
\end{equation}
Based on Lemma \ref{lem10} and (\ref{g2}), we have
\begin{eqnarray}\label{r3}
|x(t)|_{\mathcal {L}(y(t)}&\leq& \hat{c}^{\lfloor \frac{t}{T^\ast}\rfloor}
|x^0|_{\mathcal {L}(y^0)}+(4n+1)\sqrt{2}
\sum_{j=1}^{\lfloor \frac{t}{T^\ast}\rfloor}\int_{(j-1)T_\ast}^{jT_\ast}
\hat{c}^{\lfloor \frac{t}{T^\ast}\rfloor-j}|z(s)|ds+\sqrt{2}\int_{\lfloor \frac{t}{T^\ast}\rfloor}^{t}|z(s)|ds\nonumber\\
&\leq& \hat{c}^{\lfloor \frac{t}{T^\ast}\rfloor}
|x^0|_{\mathcal {L}(y^0)}+(4n+1)\sqrt{2}\int_{0}^{t}\hat{c}^{\lfloor \frac{t}{T^\ast}\rfloor-p(s)}|z(s)|ds
\end{eqnarray}
where
\begin{equation}p(s)=
\left \{
\begin{array}{ll}
i, \quad \quad s\in[(i-1)T^\ast,iT^\ast)\ for\  i=1,\dots,\lfloor \frac{t}{T^\ast}\rfloor\\
\lfloor \frac{t}{T^\ast}\rfloor, \quad \quad s\in[T^\ast\cdot\lfloor \frac{t}{T^\ast}\rfloor,t)
\end{array}
\right.
\end{equation}
Hence, (\ref{r2}) holds with $\gamma(s)=(4n+1)\sqrt{2}s$ since
$|\hat{c}^{\lfloor \frac{t}{T^\ast}\rfloor-p(s)}|\leq 1$, which
completes the proof.\hfill$\square$

\noindent{\bf Proof of Theorem \ref{thm10}}: The ``only if" part is
quite obvious, so we focus on the ``if" part.

Since $\mathcal {G}_{\sigma(t)}$ is JLC, there exists a sequence of
time instants
\begin{equation}
\label{t1} 0=T_1<T_2<\dots<T_i<T_{i+1}<\dots
\end{equation}
such that
\begin{equation}
\label{ti} T_i\triangleq
T_{i_1}<T_{i_2}<\dots<T_{i_{n+1}}=T_{i+1},\;i=1,2,\dots
\end{equation}
and  $\mathcal {G}([T_{i_{\kappa}},T_{i_{\kappa+1}}))$ is
L-connected for $\kappa=1,\dots,n$. Moreover, each arc in $\mathcal
{G}([T_{i_{k}},T_{i_{k+1}}))$ will be kept for at least the dwell
time $\tau_D$ during the time interval
$[T_{i_{\kappa}},T_{i_{\kappa+1}}),i=1,2,\dots;\kappa=1,2,\dots,n$.

Then we estimate $\Psi(t)$ during $t\in[T_i,T_{i+1}]$.  Since
$\mathcal {G}([T_{i_{1}},T_{i_{2}}))$ is L-connected, there is a
time interval $[t_1,t_1+\tau_D)\subseteq[T_{i_{1}},T_{i_{2}})$ such
that there is an edge $(l,m_0)\in\mathcal
{E}_{\sigma(t)}$ between a leader $l\in\mathcal {V}_L$ and a
follower $m_0\in\mathcal {V}_F$ for $t\in[t_1,t_1+\tau_D)$.
Based on Lemma \ref{lem11},
$$
|x_{m_0}(t_1+\tau_D)|_{\mathcal {L}(y(t_1+\tau_D))}\leq
\delta_1|x(t_1)|_{\mathcal
{L}(y(t_1))}+2\sqrt{2}\int_{t_1}^{t_1+\tau_D}|z(s)|ds,
$$
where $\delta_1\triangleq\delta(\tau_D)$.

Furthermore, we define $\mathcal {V}_L^1\triangleq \{v_{m_0}\}\cup\mathcal {V}_L$,
$$
t_2\triangleq \inf_{t}\{t\in[t_1+\tau_D,T_{i_3})|\mbox{there is an
edge leaving from}\ \mathcal {V}_L^1\ \mbox{entering}\ \mathcal
{V}\setminus \mathcal {V}_L^1\ \mbox{in}\  \mathcal
{G}_{\sigma(t)}\},
$$
and $ \mathcal {V}_F^1\triangleq\{\jmath\in\mathcal {V}_F\setminus
m_0|\mbox{there is an edge leaving from}\ \mathcal {V}_L^1\
\mbox{entering}\ \jmath\ \mbox{when}\ t=t_2\}. $

Noting that $\mathcal {G}([T_{i_{2}},T_{i_{3}}))$ is L-connected,
thus, according to Lemma \ref{lem14}, one has
$$
|x_{m_0}(t_2)|_{\mathcal {L}(y(t_2))}\leq
\delta_1|x(t_1)|_{\mathcal {L}(y(t_1))}+\sqrt{2}
\int_{t_1+\tau_D}^{t_2}|z(s)|ds+2\sqrt{2}\int_{t_1}^{t_1+\tau_D}|z(s)|ds
$$
Further, by Lemma \ref{lem13},
\begin{eqnarray}\label{41}
|x_{m_0}(t)|_{\mathcal {L}(y(t))}&\leq &
\phi_{1}|x(t_1)|_{\mathcal
{L}(y(t_1))}+\sqrt{2}[2\int_{t_1}^{t_2+\tau_D}+\int_{t_1+\tau_D}^{t_2}+2\int_{t_1}^{t_1+\tau_D}]|z(s)|ds\nonumber\\
&\leq &\phi_{1}|x(t_1)|_{\mathcal
{L}(y(t_1))}+4\sqrt{2}\int_{t_1}^{t_2+\tau_D}|z(s)|ds
\end{eqnarray}
for $t\in[t_2,t_2+\tau_D]$, where
$\phi_{1}\triangleq\phi_{\delta_1}(\tau_D)$. Moreover, according to
Lemma \ref{lem12},
\begin{eqnarray}\label{42}
|x_{\jmath}(t_2+\tau_D)|_{\mathcal {L}(y(t_2+\tau_D))} &\leq &
\varphi_{\phi_1}(\tau_D)|x(t_1)|_{\mathcal
{L}(y(t_1))}+\sqrt{2}[4\int_{t_1}^{t_2+\tau_D}
+2\int_{t_2}^{t_2+\tau_D}+2\int_{t_1}^{t_2}]|z(s)|ds\nonumber\\
&\leq& \varphi_{\phi_1}(\tau_D)|x(t_1)|_{\mathcal {L}(y(t_1))} +8\sqrt{2}
\int_{t_1}^{t_2+\tau_D}|z(s)|ds
\end{eqnarray}
for $\jmath\in \mathcal {V}^1_F$. Because
$\phi_{1}<\varphi_{\phi_1}(\tau_D)$,  (\ref{41}) and (\ref{42}) lead
to
$$
|x_i(t_2+\tau_D)|_{\mathcal {L}(y(t_2+\tau_D))}\leq
\delta_2|x(t_1)|_{\mathcal
{L}(y(t_1))}+8\sqrt{2}\int_{t_1}^{t_2+\tau_D}|z(s)|ds,\; \forall i\in
\{v_{m_0}\}\cup\mathcal {V}^1_F,
$$
where $\delta_2\triangleq\varphi_{\phi_1}(\tau_D)$.

Next, define $\mathcal {V}^2_L\triangleq \mathcal {V}^1_L\cup\mathcal
{V}^1_F$,
$$
t_3\triangleq \inf_{t}\{t\in[t_2+\tau_D,T_{i_4})|\mbox{there is an
edge leaving from}\ \mathcal {V}^2_L\ \mbox{and entering}\ \mathcal
{V}\setminus\mathcal {V}_L^2\ \mbox{in}\  \mathcal {G}_{\sigma(t)}\}
$$
and $ \mathcal {V}^2_F\triangleq\{\jmath\in\mathcal
{V}\setminus\mathcal {V}_L^2|\mbox{there is an edge leaving from}\
\mathcal {V}^2_L\ \mbox{entering}\ \jmath\ \mbox{when}\ t=t_3\}. $

Similarly, from Lemma \ref{lem14}, by $
\phi_{2}\triangleq\phi_{\delta_2}(\tau_D),\;\delta_3\triangleq\varphi_{\phi_2}(\tau_D),
$ one has
$$
|x_i(t_3+\tau_D)|_{\mathcal {L}(y(t_3+\tau_D))}\leq
\delta_3|x(t_1)|_{\mathcal
{L}(y(t_1))}+12\sqrt{2}\int_{t_1}^{t_3+\tau_D}|z(s)|ds, \quad \forall i\in
\{v_{m_0}\}\cup\mathcal {V}^1_F\cup\mathcal {V}^2_F
$$
Repeating the process gives
$$
\phi_{\kappa}\triangleq\phi_{\delta_\kappa}(\tau_D),\quad\delta_{\kappa+1}\triangleq\varphi_{\phi_\kappa}(\tau_D),
$$
for $\kappa=3,4,\dots,k_0$ until $\mathcal
{V}_F=\{v_{m_0}\}\cup\mathcal {V}^1_F\cup \mathcal
{V}^2_F\cup\dots\cup\mathcal {V}^{k_0}_F$ for some $k_0\leq n$ such
that
$$
|x_i(t_{k_0}+\tau_D)|_{\mathcal {L}(y(t_{k_0}+\tau_D))}\leq
\delta_{k_0}|x(t_1)|_{\mathcal
{L}(y(t_1))}+4\sqrt{2}k_0\int_{t_1}^{t_{k_0}+\tau_D}|z(s)|ds, \quad
\forall i \in\mathcal {V}_F
$$
Hence
$$
|x(t_{k_0}+\tau_D)|_{\mathcal {L}(y(t_{k_0}+\tau_D))}\leq
\delta_{k_0}|x(t_1)|_{\mathcal
{L}(y(t_1))}+4\sqrt{2}k_0\int_{t_1}^{t_{k_0}+\tau_D}|z(s)|ds
$$
According to Lemma \ref{lem14}, we obtain
$$
|x(T_{i+1})|_{\mathcal {L}(y(T_{i+1}))}\leq
\delta_{k_0}|x(T_i)|_{\mathcal
{L}(y(T_{i}))}+(4k_0+1)\sqrt{2}\int_{T_{i}}^{T_{i+1}}|z(s)|ds.
$$
It is obvious to see that $k_0\leq n$ and $0<\delta_{1}\leq \delta_{2}\leq\dots \delta_{n}
<1$. Therefore, denote $\hat{\delta}\triangleq  \delta_{n}$, then for $K=1,2,\dots$,
\begin{equation}\label{120}
|x(T_{K+1})|_{\mathcal {L}(y(T_{K+1}))}\leq
\hat{\delta}^{K}|x^0|_{\mathcal
{L}(y^0)}+(4n+1)\sqrt{2}\sum_{i=1}^{K} \hat{\delta}^{K-i} \int_{T_{i}}^{T_{i+1}}|z(s)|ds
\end{equation}
Thus, similar to the proof of Theorem \ref{thmr1}, we also have
\begin{eqnarray}\label{r4}
|x(t)|_{\mathcal {L}(y(t)}
\leq \hat{\delta}^{\Gamma(t)}
|x^0|_{\mathcal {L}(y^0)}+(4n+1)\sqrt{2}\int_{0}^{t}\hat{\delta}^{\Gamma(t)-\hat{p}(s)}|z(s)|ds
\end{eqnarray}
where $\Gamma(t)=K_0-1$ when
$t\in[T_{K_0},T_{K_0+1}),K_0=1,2,\dots$, and
\begin{equation}\hat{p}(s)=
\left \{
\begin{array}{ll}
i, \quad \quad s\in[T_i,T_{i+1})\  for\  i=1,\dots,K_0-1\\
K_0-1, \quad \quad s\in[T_{K_0},t)
\end{array}
\right.
\end{equation}
Then it is obvious to see that (\ref{r4}) leads to Theorem
\ref{thm10} immediately.\hfill$\square$

\noindent{\bf Proof of Theorem \ref{thm12}}: We also focus on the
``if" part since the ``only if" part is quite obvious.

Because $\mathcal {G}_{\sigma(t)}$ is JLC, there is an infinite
sequence in the form of (\ref{t1}) with (\ref{ti}) such that
$\mathcal {G}([T_{i_{\kappa}},T_{i_{\kappa+1}}))$ is L-connected for
$\kappa=1,\dots,n$.

Then, for any $\ell \in \mathcal {V}_1$, there is
$t_\ell\in[T_{i_1},T_{i_2})$ such that there is an arc leaving from
$\mathcal {V}_F$ entering $\ell$ in $\mathcal {G}_{\sigma(t)}$.
Hence, recalling Lemma \ref{lem11},
$$
|x_{\ell}(t_\ell+\tau_D)|_{\mathcal {L}(y(t_\ell+\tau_D))}\leq d_1
|x(t_\ell)|_{\mathcal
{L}(y(t_\ell))}+2\sqrt{2}\int_{t_\ell}^{t_\ell+\tau_D}|z(s)|ds, \;\ell \in \mathcal
{V}_1^F
$$
with a constant $d_1\triangleq\delta(\tau_D)$. According to Lemma
\ref{claim}, for any $ \ell \in \mathcal {V}_1^F$, we have
$$
|x_{\ell}(t)|_{\mathcal {L}(y(t))}\leq d_1|x(T_i)|_{\mathcal
{L}(y(T_i))} +2\sqrt{2}\int_{T_i}^{T_{i+1}}|z(s)|ds,\; t\in
[T_{i_2},T_{i+1}]
$$
Again by Lemmas \ref{lem12}
and \ref{claim}, for any $ \ell \in \mathcal {V}_2^F$,
$$
|x_{\ell}(t)|_{\mathcal {L}(y(t))} \leq d_2|x(T_i)|_{\mathcal
{L}(y(T_i))}+4\sqrt{2}\int_{T_i}^{T_{i+1}}|z(s)|ds ,\;t\in
[T_{i_3},T_{i+1}],
$$
where $d_2=\varphi_{d_1}(\tau_D)$. Similarly, with
$d_j=\varphi_{d_{j-1}}(\tau_D),j=3,\dots,k_0$,
$$
|x_{\ell}(t)|_{\mathcal {L}(y(t))} \leq d_j|x(T_i)|_{\mathcal
{L}(y(T_i))}+2\sqrt{2}j\int_{T_i}^{T_{i+1}}|z(s)|ds ,\; t\in
[T_{i_{j+1}},T_{i+1}],
$$
for any $ \ell \in \mathcal {V}_j^F,j=3,\dots,k_0$, which leads to
$$
|x(T_{i+1})|_{\mathcal {L}(y(T_{i+1}))} \leq
d_{k_0}|x(T_i)|_{\mathcal
{L}(y(T_i))}+2\sqrt{2}k_0\int_{T_i}^{T_{i+1}}|z(s)|ds.
$$
Similar to the proof of Theorem \ref{thm10}, SiISS can be
obtained.\hfill$\square$

\section{Conclusions}

This paper addressed multi-agent set tracking problems with multiple
leaders and switching communication topologies. At first, the
equivalence between UJLC and the SISS of a group of uncertain agents
with respect to a moving multi-leader set was shown.  Then it was
shown that UJLC is a sufficient condition for SiISS of the
multi-agent system with disturbances in agent dynamics and
unmeasurable velocities in the dynamics of the leaders. Moreover,
when communication topologies are either bidirectional or acyclic,
JLC is a necessary and sufficient condition for SiISS. Also, set
tracking was achieved in special cases with the help of SISS and
SiISS.

%In many practical situations, agent dynamics in a multi-agent
%network are different from each other; the inter-agent communication
%is directed, time-delayed and asynchronous; the communication links
%and sensing information are unreliable; the ranges of agent sensors
%and actuators are limited; the collective task are hybrid and
%complicated.
Multiple leaders, in some practical cases, can provide an effective
way to overcome the difficulties and constraints in the distributed
design. On the other hand, ISS-based tools were proved to be very
powerful in the control synthesis. Therefore, the study of multiple
active leaders and related ISS tools deserves more attention.

\section*{Appendix}
\noindent{\bf A.1 \; Proof of Lemma \ref{lem8}}

Due to $D^+\sqrt{\Psi(t)}\leq \sqrt{2}\|z\|_{\infty}$ by Lemma
\ref{lem10} and (\ref{g2}), we obtain
\begin{equation}\label{10}
\sqrt{\psi_j(t)}\leq \sqrt{\Psi(t)}\leq \sqrt{\Psi(t_0)}
+ \sqrt{2}\|z\|_{\infty}(t-t_0), \; j=1,...,n.
\end{equation}

Since there is an arc $(j,i)$ with $j\in\mathcal {V}_L$ and
$i\in\mathcal {V}_F$ in $\mathcal {G}_{\sigma(t)}$ for
$t\in[t_0,t_0+\tau_D)$, based on (\ref{31}), one has
\begin{eqnarray}
\langle x_i-\mathcal {P}_{\mathcal
{L}(y(t))}(x_i), \sum_{j \in
L_i(\sigma(t))}b_{ij}(x)(y_j-x_i)\rangle\leq-b_\ast
\psi_i(t).
\end{eqnarray}

Thus, with (\ref{e12}) and the fact that $r(t)+w_i(t)\leq q(t) \leq \sqrt{2}\|z\|_{\infty}$, we obtain
\begin{eqnarray}\label{13}
\frac{d}{dt}\psi_i(t)
&\leq&-2b_\ast
\psi_i(t)+2\langle
x_i-\mathcal {P}_{\mathcal {L}(y(t))}(x_i), \sum\limits_{j \in
N_i(\sigma(t))}a_{ij}(x_j-x_i)\rangle+2(r(t)+w_i(t))\sqrt{\psi_i(t)} \nonumber\\
&\leq& -2b_\ast
\psi_i(t)+2\langle
x_i-\mathcal {P}_{\mathcal {L}(y(t))}(x_i), \sum\limits_{j \in
N_i(\sigma(t))}a_{ij}(x_j-x_i)\rangle+2\sqrt{2}\|z\|_{\infty} \sqrt{\psi_i(t)}
\end{eqnarray}
for  $t\in [t_0,t_0+\tau_D)$.

Then, by Lemma \ref{lem6}, if $\sqrt{\psi_j(t)}<\sqrt{\psi_i(t)}, j \in
N_i(\sigma(t))$ for $t\in [t_0,t_0+\tau_D)$, then
\begin{equation}\label{r12}
\langle x_i(t)-\mathcal {P}_{\mathcal {L}(y(t))}(x_i(t)),x_j(t)-x_i(t)\rangle \leq 0.
\end{equation}
On the other hand, if $\sqrt{\psi_j(t)}\geq\sqrt{\psi_i(t)}, j \in
N_i(\sigma(t))$, from Lemma \ref{lem6} and (\ref{10}),
\begin{eqnarray}\label{11}
\langle x_i(t)-\mathcal {P}_{\mathcal {L}(y(t))}(x_i(t)),x_j(t)-x_i(t)\rangle
&\leq& \sqrt{\psi_i(t)}(\sqrt{\psi_j(t)}-\sqrt{\psi_i(t)})\nonumber\\
&\leq& \sqrt{\psi_i(t)}(\sqrt{\Psi(t_0)} +\sqrt{2}\|z\|_{\infty}(t-t_0)-\sqrt{\psi_i(t)})\nonumber\\
&\leq& \sqrt{\psi_i(t)}(\sqrt{\Psi(t_0)}-\sqrt{\psi_i(t)}
+\sqrt{2}\|z\|_{\infty}\tau_D)
\end{eqnarray}
$t\in [t_0,t_0+\tau_D)$. Therefore, with (\ref{13}), (\ref{r12}) and (\ref{11}), it follows that
$$
\frac{d}{dt}\psi_i(t) \leq
-2\lambda\psi_i(t)+2[\sqrt{2}\|z\|_{\infty}(1+(n-1)a^\ast\tau_D)
+(n-1)a^\ast\sqrt{\Psi(t_0)}]\sqrt{\psi_i(t)},
$$
where $\lambda\triangleq b_\ast+(n-1)a^\ast$, or equivalently,
$$
\frac{d}{dt}\sqrt{\psi_i(t)}
\leq-\lambda\sqrt{\psi_i(t)}+[\sqrt{2}\|z\|_{\infty}
(1+(n-1)a^\ast\tau_D)+(n-1)a^\ast\sqrt{\Psi(t_0)})]
$$
for $t\in [t_0,t_0+\tau_D)$.  As a result,
\begin{eqnarray}\label{14}
\sqrt{\psi_i(t)} &\leq& e^{-\lambda(t-t_0)}\sqrt{\Psi(t_0)}+
(1-e^{-\lambda(t-t_0)})\frac{(n-1)a^\ast\sqrt{\Psi(t_0)}+(\|u\|_{\infty}
+\|w\|_{\infty})(1+(n-1)a^\ast\tau_D)}{\lambda}\nonumber\\
%&=& \frac{b_\ast
%e^{-\lambda(t-t_0)}+(n-1)a^\ast}{b_\ast+(n-1)a^\ast}\sqrt{\Psi(t_0)}
%+(1-e^{-\lambda(t-t_0)})\sqrt{2}\|z\|_{\infty}
%\frac{1+(n-1)a^\ast\tau_D}{\lambda}\nonumber\\
&\leq&  \hat{\mu}(t-t_0)
\sqrt{\Psi(t_0)}+c_0\sqrt{2}\|z\|_{\infty},\qquad\qquad t\in
[t_0,t_0+\tau_D)
\end{eqnarray}
where $\hat{\mu}(s)\triangleq \frac{b_\ast e^{-\lambda
s}+(n-1)a^\ast}{b_\ast+(n-1)a^\ast},\, s\in[0,\tau_D]$ and
$c_0\triangleq \frac{1+(n-1)a^\ast\tau_D}{b_\ast+(n-1)a^\ast}$,
because $1-e^{-\lambda(t-t_0)}<1$.

Then we evaluate $\sqrt{\psi_i(t)}$ for $t\in
[t_0+\tau_D,t_0+T_\ast)$ no matter whether there is any connection
between the followers and the leaders. Similar analysis gives
\begin{eqnarray}
\frac{d}{dt}\psi_i(t) &\leq& 2\sqrt{2}\|z\|_{\infty}
\sqrt{\psi_i(t)}+2\langle x_i-\mathcal {P}_{\mathcal {L}(y(t))}(x_i),
\sum\limits_{j \in N_i(\sigma(t))}a_{ij}(x)(x_j-x_i)\rangle\nonumber\\
&\leq& 2\sqrt{2}\|z\|_{\infty} \sqrt{\psi_i(t)}+2(n-1)
a^\ast\sqrt{\psi_i(t)}(\sqrt{\Psi(t_0)}-\sqrt{\psi_i(t)} +\sqrt{2}\|z\|_{\infty} T_\ast)\nonumber\\
&=&
-2(n-1)a^\ast\psi_i(t)+2[\sqrt{2}\|z\|_{\infty}(1+(n-1)a^\ast
T_\ast)+(n-1)a^\ast\sqrt{\Psi(t_0)}]\sqrt{\psi_i(t)},\nonumber
\end{eqnarray}
which is equivalent to
\begin{eqnarray}
\frac{d}{dt}\sqrt{\psi_i(t)}
\leq-(n-1)a^\ast\sqrt{\psi_i(t)}+[\sqrt{2}\|z\|_{\infty}(1+(n-1)a^\ast
T_\ast)+(n-1)a^\ast\sqrt{\Psi(t_0)}].
\end{eqnarray}
Denote $\hat{\mu}^\ast\triangleq \hat{\mu}(\tau_D) $.  From (\ref{14}),
when $t\in [t_0+\tau_D,t_0+T_\ast)$,
\begin{eqnarray}
\sqrt{\psi_i(t)}\label{15}
&\leq& e^{-(n-1)a^\ast(t-(t_0+\tau_D))}\sqrt{\psi_i(t_0+\tau_D)}\nonumber\\
&&+(1-e^{-(n-1)a^\ast(t-(t_0+\tau_D))})[\sqrt{\Psi(t_0)}+\sqrt{2}\|z\|_{\infty}
\frac{1+(n-1)a^\ast T_\ast}{(n-1)a^\ast}]\nonumber\\
&\leq& e^{-(n-1)a^\ast(t-(t_0+\tau_D))}[\hat{\mu}^\ast
\sqrt{\Psi(t_0)}+c_0\sqrt{2}\|z\|_{\infty}]\nonumber\\
&&+(1-e^{-(n-1)a^\ast(t-(t_0+\tau_D))})[\sqrt{\Psi(t_0)}+\sqrt{2}\|z\|_{\infty}\cdot\frac{1+(n-1)a^\ast
T_\ast}{(n-1)a^\ast}]\nonumber\\&\leq& \tilde{\mu} (t-t_0)
\sqrt{\Psi(t_0)} + \gamma_1\|z\|_{\infty},
\end{eqnarray}
where $\gamma_1\triangleq \sqrt{2}\cdot \frac{1+(n-1)a^\ast
T_\ast}{(n-1)a^\ast}>c_0$ and $\tilde{\mu}  (s)\triangleq
1-e^{-(n-1)a^\ast(s-\tau_D))}(1-\hat{\mu}^\ast),
s\in[\tau_D,T_\ast]$. Therefore, based on (\ref{14}) and (\ref{15}),
$$
\sqrt{\psi_i(t)}\leq
\mu(t-t_0)\sqrt{\Psi(t_0)}+\gamma_1\|z\|_{\infty},\quad
\mu(s)=\left\{
\begin{array}{ll}
\hat{\mu} (s),&\mbox{$s\in [0,\tau_D)$}\\
\tilde{\mu}(s),&\mbox{$s\in [\tau_D,T_\ast]$}
\end{array}
\right.
$$
where $\mu(s)$ is continuous. Thus, the conclusion
follows.\hfill$\square$

\noindent {\bf A.2 \;Proof of Lemma \ref{lem9}}

 If there is an arc $(v_i,v_m)$ in $\mathcal
{G}_{\sigma(t)}$ for $t\in[t_0,t_0+\tau_D)$, then based on
(\ref{10}), Lemmas \ref{lem6} and \ref{lem7}, it is easy to see
\begin{eqnarray}
\frac{d}{dt}\psi_m(t)&\leq&2\langle x_m-\mathcal {P}_{\mathcal
{L}(y(t))}(x_m), \sum\limits_{j \in
N_m(\sigma(t))}a_{mj}(x)(x_j-x_m)+\sum_{j \in
L_m(\sigma(t))}b_{mj}(x)(y_j-x_m)\rangle\nonumber\\
&&+2\sqrt{2}\|z\|_{\infty} \sqrt{\psi_m(t)}\nonumber\\
&\leq& 2\sqrt{2}\|z\|_{\infty} \sqrt{\psi_m(t)}+2\langle
x_m-\mathcal {P}_{\mathcal {L}(y(t))}(x_m),
\sum\limits_{j \in N_m(\sigma(t))}a_{mj}(x)(x_j-x_m)\rangle\nonumber\\
&=& 2\sqrt{2}\|z\|_{\infty} \sqrt{\psi_m(t)}+2\sum\limits_{j \in N_m(\sigma(t))\setminus v_i}a_{mj}(x)\langle
x_m-\mathcal {P}_{\mathcal {L}(y(t))}(x_m),
x_j-x_m\rangle\nonumber\\
&&+2a_{mi}(x)\langle x_m-\mathcal {P}_{\mathcal {L}(y(t))}(x_m),
x_i-x_m\rangle\nonumber\\
&\leq& 2\sqrt{2}\|z\|_{\infty} \sqrt{\psi_m(t)}+2(n-2)a^\ast\sqrt{\psi_m(t)}
(\sqrt{\Psi(t_0)}-\sqrt{\psi_m(t)} +\sqrt{2}\|z\|_{\infty}\tau_D)\nonumber\\
&&- 2a_\ast
\sqrt{\psi_m(t)}(\sqrt{\psi_m(t)}-\sqrt{\psi_i(t)})\nonumber
\end{eqnarray}
for $t\in [t_0,t_0+\tau_D)$. Then, if (\ref{16}) holds, as done in
the proof of Lemma \ref{lem8}, we can obtain
$$
\frac{d}{dt}\sqrt{\psi_m(t)} \leq -\lambda_1\sqrt{\psi_m(t)}+\hat
d_1
$$
where $\lambda_1\triangleq (n-2)a^\ast+a_\ast$ and $\hat d_1
\triangleq
(1+(n-2)a^\ast\tau_D)\sqrt{2}\|z\|_{\infty}+((n-2)a^\ast+a_\ast\mu_0)\sqrt{\Psi(t_0)}+a_\ast
d_0$.  Here are two cases.
\begin{itemize}
\item when $t\in [t_0,t_0+\tau_D)$:
\begin{eqnarray}\label{28}
\sqrt{\psi_m(t)} &\leq& e^{-\lambda_1(t-t_0)}\sqrt{\psi_m(t_0)}+
(1-e^{-\lambda_1(t-t_0)})\frac{\hat
d_1}{\lambda_1}\nonumber\\
&\leq&
\frac{(n-2)a^\ast+(\mu_0+(1-\mu_0)e^{-\lambda_1(t-t_0)})a_\ast }
{(n-2)a^\ast+a_\ast}\sqrt{\Psi(t_0)}+(1-e^{-\lambda_1(t-t_0)})\nonumber\\
&&\cdot\frac{\sqrt{2}\|z\|_{\infty}(1+(n-2)a^\ast\tau_D)+a_\ast d_0}{(n-2)a^\ast+a_\ast}\nonumber\\
&\leq&  \hat{\xi}(t-t_0)
\sqrt{\Psi(t_0)}+\gamma_0,
\end{eqnarray}
where $\hat{\xi}(s)\triangleq
\frac{(n-2)a^\ast+(\mu_0+(1-\mu_0)e^{-\lambda_1 s})a_\ast
}{(n-2)a^\ast+a_\ast},\,s\in [0,\tau_D]$ and $\gamma_0\triangleq
\frac{(1+(n-2)a^\ast\tau_D)\sqrt{2}\|z\|_{\infty}+a_\ast d_0}{(n-2)a^\ast+a_\ast}$.

\item when $t\in [t_0+\tau_D,t_0+T_\ast)$: Denote
$\hat{\xi}^\ast\triangleq \hat{\xi}(\tau_D) $. By (\ref{28}),
similarly, we have
\begin{eqnarray}
\sqrt{\psi_m(t)}
&\leq& e^{-(n-1)a^\ast(t-(t_0+\tau_D))}\sqrt{\psi_m(t_0+\tau_D)}\nonumber\\
&&+(1-e^{-(n-1)a^\ast(t-(t_0+\tau_D))})[\sqrt{\Psi(t_0)}+
\sqrt{2}\|z\|_{\infty}\frac{1+(n-1)a^\ast T_\ast}{(n-1)a^\ast}]\nonumber\\
&\leq& e^{-(n-1)a^\ast(t-(t_0+\tau_D))}[\hat{\xi}^\ast \sqrt{\Psi(t_0)}+\gamma_0]\nonumber\\
&&+(1-e^{-(n-1)a^\ast(t-(t_0+\tau_D))})[\sqrt{\Psi(t_0)}+\sqrt{2}\|z\|_{\infty}\frac{1+(n-1)a^\ast
T_\ast}{(n-1)a^\ast}]\nonumber\\&\leq& \tilde{\xi} (t-t_0)
\sqrt{\Psi(t_0)} + \gamma_2\|z\|_{\infty}+d_0, \label{24}
\end{eqnarray}
where $ \gamma_2 \triangleq \sqrt{2}\cdot\frac{1+(n-1)a^\ast
T_\ast}{(n-2)a^\ast+a_\ast} $ and $\tilde{\xi} (s)\triangleq
1-e^{-(n-1)a^\ast(s-\tau_D)}(1-\hat{\xi} ^\ast) ,\;
s\in[\tau_D,T_\ast]$, because
$$
\max\{\gamma_0,\sqrt{2}\|z\|_{\infty}\frac{1+(n-1)a^\ast
T_\ast}{(n-1)a^\ast}\}\leq \sqrt{2}\|z\|_{\infty}
\frac{1+(n-1)a^\ast T_\ast}{(n-2)a^\ast+a_\ast}+d_0,
$$
\end{itemize}
With (\ref{28}) and (\ref{24}), we have
$$
\sqrt{\psi_m(t)}\leq
\xi_{\mu_0}(t-t_0)\sqrt{\Psi(t_0)}+\gamma_2\|z\|_{\infty}+d_0,\quad
\xi_{\mu_0}(s)=\left\{
\begin{array}{ll}
\hat{\xi}  (s),&\mbox{$s\in [0,\tau_D)$}\\
\tilde{\xi}  (s),&\mbox{$s\in [\tau_D,T_\ast]$}
\end{array}
\right.
$$
where $\xi_{\mu_0}(s)$ which is continuous. Thus, the conclusion
follows. \hfill$\square$

\end{document}